\documentclass[12pt]{article}

\RequirePackage{amsthm,amsmath,amsfonts,amssymb}
\RequirePackage[]{natbib}
\RequirePackage[colorlinks,citecolor=blue,urlcolor=blue]{hyperref}
\RequirePackage{graphicx}

\theoremstyle{plain}

\newtheorem{theorem}{Theorem}[section]

\theoremstyle{remark}
\newtheorem{definition}[theorem]{Definition}

\usepackage{graphicx}
\usepackage{enumerate}
\usepackage{natbib}
\usepackage{url} 
\usepackage{breqn}
\usepackage{longtable}
\usepackage{amssymb}
\usepackage[dvips]{epsfig}
\usepackage{dsfont}
\usepackage[usenames,dvipsnames]{xcolor}
\usepackage{rotating}
\usepackage{makecell}
\usepackage[percent]{overpic}
\usepackage[font={stretch=1.0,footnotesize}]{caption}
\usepackage{mathtools}
\usepackage{algorithmicx}
\usepackage[]{algpseudocode}
\usepackage[section]{algorithm}
\usepackage{diagbox}
\usepackage{wrapfig}
\input epsf
\usepackage{fontenc}
\usepackage{bm}
\usepackage{slashbox}
\usepackage{multirow}
\usepackage{eurosym}
\usepackage{bbm}
\usepackage{longtable}
\usepackage[flushleft]{threeparttable}
\epsfverbosetrue

\usepackage{titlesec}

\usepackage{scalerel,stackengine}
\stackMath
\newcommand\reallywidehat[1]{%
\savestack{\tmpbox}{\stretchto{%
  \scaleto{%
    \scalerel*[\widthof{\ensuremath{#1}}]{\kern-.6pt\bigwedge\kern-.6pt}%
    {\rule[-\textheight/2]{1ex}{\textheight}}
  }{\textheight}%
}{0.5ex}}%
\stackon[1pt]{#1}{\tmpbox}%
}
\usepackage{paralist}

\renewenvironment{itemize}[1]{\begin{compactitem}#1}{\end{compactitem}}

\def \dsP {\text{$\mathds{P}$}}
\def \dsE {\text{$\mathds{E}$}}
\def \dsR {\text{$\mathds{R}$}}

\DeclareMathOperator{\Var}{Var}

\DeclareMathOperator{\diag}{diag}

\DeclareMathOperator{\SDD}{SD}

\DeclareMathOperator{\ND}{N}

\def \qlt         {\text{$q_{\lambda}(\vartheta)$}}

\begin{document}


\begin{titlepage}
\title{
Actor Heterogeneity and Explained Variance in Network Models---A Scalable Approach through Variational Approximations}

\author{Nadja Klein$^{1,\ast}$ and G{\"o}ran Kauermann$^2$ \\
\small
$^1$Technische Universit\"at Dortmund,
$^2$Ludwig-Maximlians-Unversit\"at M\"unchen}
\vspace{6cm}
\date{}
\maketitle
\noindent

\begin{abstract}
The analysis of network data has gained considerable interest in recent years. This also includes the analysis of large, high-dimensional networks with hundreds and thousands of nodes. While exponential random graph models serve as workhorse for network data analyses, their applicability to very large networks is problematic via classical inference such as maximum likelihood or exact Bayesian estimation owing to scaling and instability issues. The latter trace from the fact that classical network statistics consider nodes as exchangeable, i.e.,~actors in the network are assumed to be homogeneous. This is often questionable. One way to circumvent the restrictive assumption is to include actor-specific random effects, which account for unobservable heterogeneity. However, this increases the number of unknowns considerably, thus making the model highly-parameterized. As a solution even for very large networks, we propose a scalable approach based on variational approximations, which not only leads to numerically stable estimation but is also applicable to high-dimensional directed as well as undirected networks. We furthermore demonstrate that including node-specific covariates can reduce node heterogeneity, which we facilitate through versatile prior formulations and a new measure that we call \textit{posterior explained variance}. We illustrate our approach in three diverse examples, covering network data from the Italian Parliament, international arms trading, and Facebook; and conduct detailed simulation studies.
\end{abstract}

\noindent
{\bf Keywords}: Exponential random graph model; node heterogeneity; random effects; re-parameterization trick; social network analysis; stochastic variational inference

\vfill
{\noindent\small 
$\mbox{}^\star$ Correspondence should be directed to Nadja Klein at nadja.klein@tu-dortmund.de.}

\end{titlepage}


\section{Introduction}\label{sec:intro}
Until about two decades ago, the analysis of network data was relatively rare in the literature. However, it has since become a continuously emerging field in statistics, allowing the investigation of 
links and relationships between nodes representing, for instance, social entities, genes, diseases, and many others 
\citep[see][for examples of overviews on this topic]{GolZheFie2009,Kol2009,Fie2012,KolCsa2014,BiaKauMey2019}. 

A key statistical model for network data analyses is the exponential random graph model (ERGM),  originally proposed by \citet{FraStr1986}. ERGMs may certainly be considered as the workhorses in statistical network data analyses, as they allow for interpretable models as well as extensions in various ways. Nowadays, ERGMs are widely used
in economics, sociology, and political science, amongst others, to
analyse (social) networks. The key idea is to consider a network with $N$ nodes as a binary adjacency matrix $Y \in \{0,1\}^{N\times N}$, where  $Y_{ij}= 1 $ if an edge from node $i$ to node $j$ exists, and $Y_{ij}= 0 $ otherwise. The diagonal is  left undefined, i.e.,~edges connecting
a node to itself are not allowed, and we set $Y_{ii} = 0$.  The network may be directed or undirected, depending on the relationships between the actors in the network. Compared to directed networks, undirected networks fulfil the symmetry property that $Y_{ij} = Y_{ji}$ for all $1 \le i < j \le N$. The adjacency matrix $Y$ can be considered as a random variable with a probability distribution that is based on a set of sufficient network statistics $s_k(y)$, for $k = 0, \ldots p$. This leads to the probability model
\begin{align}
\label{eq:ergm}
\dsP(Y=y\mid\beta) &= \frac{ \exp( s(y)^\top \beta) }{\kappa(\beta)},
\end{align}
where $s(\cdot) = (s_0(\cdot), s_1(\cdot),\ldots ,  s_{p}(\cdot))^\top$ is a vector of  sufficient network statistics, such as the number of edges or two-stars in a network \citep[see e.g.][ and Supplement~A for relevant network statistics]{SniPatRobHan2006} and $\beta = (\beta_0, \beta_1, \ldots , \beta_p)^\top\in\dsR^{p+1}$ is the  parameter vector of model coefficients. Since  $s(y)$ depends on the network only, its components are endogenous.  The model in \eqref{eq:ergm} represents an exponential family distribution, and the  interpretation of an ERGM and its coefficients $\beta$  is similar to that of a binary logistic regression model. That is, a tie in a network is the outcome, and the characteristics of network members  and network structures aid in explaining or predicting the probability of a tie~\citep{HunGooHan2008}. However, unlike in logistic regression, the normalization constant $\kappa(\beta)=\sum_{y^\star \in \mathcal{Y}} \exp(s({y}^\star)^\top{\beta})$ involves a sum over all possible networks represented by $\mathcal{Y}$ being the set of all (symmetric for undirected and non-symmetric for directed networks) $\{0, 1\}^{N \times N}$ matrices. 
This is intractable except for very small networks. Estimation of $\kappa(\beta)$ in  model \eqref{eq:ergm}  therefore needs to be carried out approximately or based on simulations \citep[see also][for a general survey on computational methods used in network analysis]{HunKriSch2012}. 

An early reference for estimating ERGMs is \citet{Sni2002}, who adapted so-called Markov Chain Monte Carlo (MCMC) maximum likelihood estimation (MLE)~\citep[MCMC-MLE][]{GeyTho1992}. A numerically stable routine thereof has been proposed in \citet{HumHunHan2012} using a so-called stepping algorithm. However,  MCMC-MLE may have exponentially slow convergence, making estimation  impractical  for a large class of ERGMs \citep[][]{ChaDia2013}.  Bayesian estimation of ERGMs is challenging as well and known as a \emph{doubly intractable} problem, since not only $\kappa(\beta)$ but also the normalizing constant of the posterior is intractable. \citet{CaiFri2011} therefore proposed an MCMC algorithm that samples from the
likelihood using a ``tie no tie'' sampler \citep{HunGooHan2008} coupled with the exchange algorithm of \citet{MurGhaMac2006} to sample from the posterior. Reducing the numerical burden of Bayesian estimation in model \eqref{eq:ergm} is also pursued by \citet{YinBut2020} or by \citet{BouFriMai2018}, who propose  adjusted pseudolikelihood estimation in the context of Bayesian model selection. 

\begin{figure*}[htbp]
\centering
\caption{Facebook egos.}\label{fig:facebook}
\includegraphics[width=0.45\textwidth]{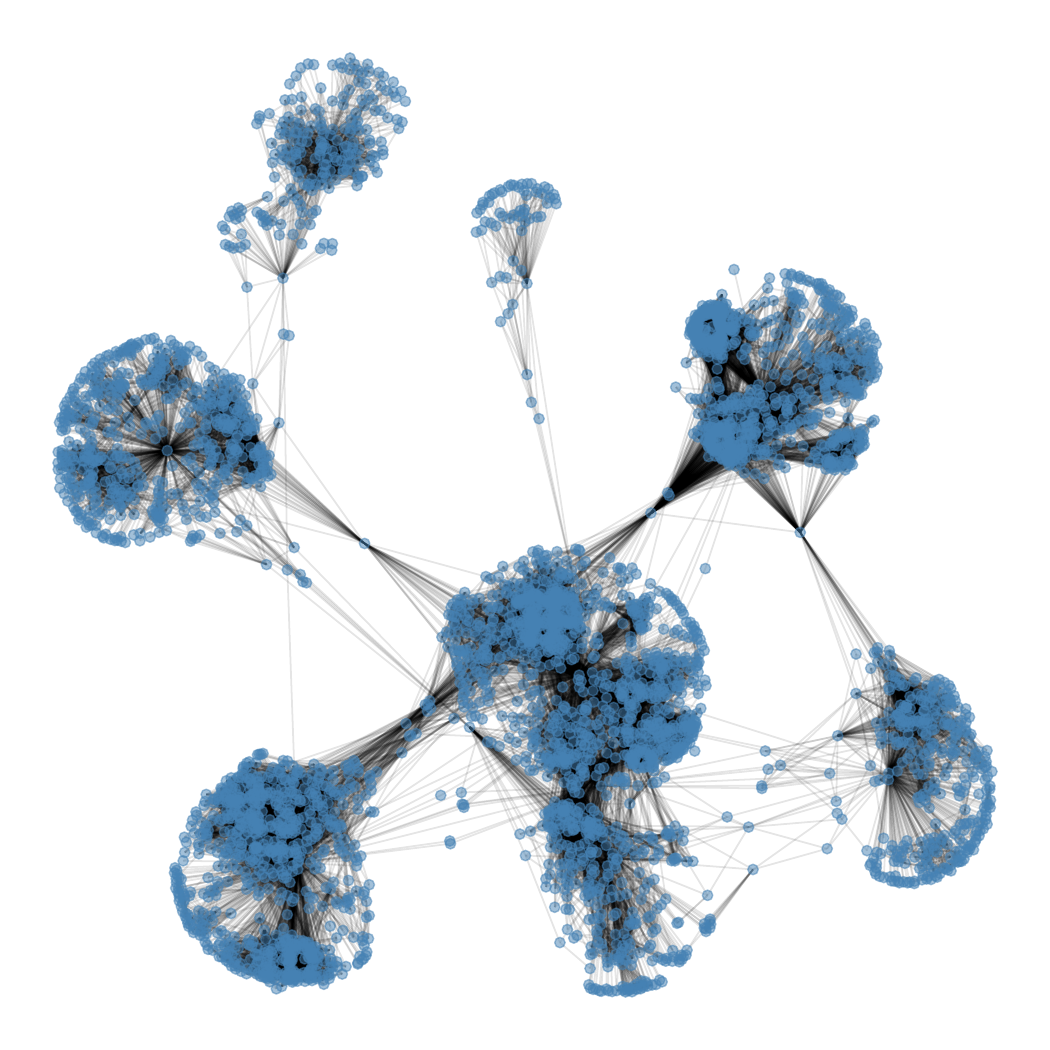}
\includegraphics[width=0.45\textwidth]{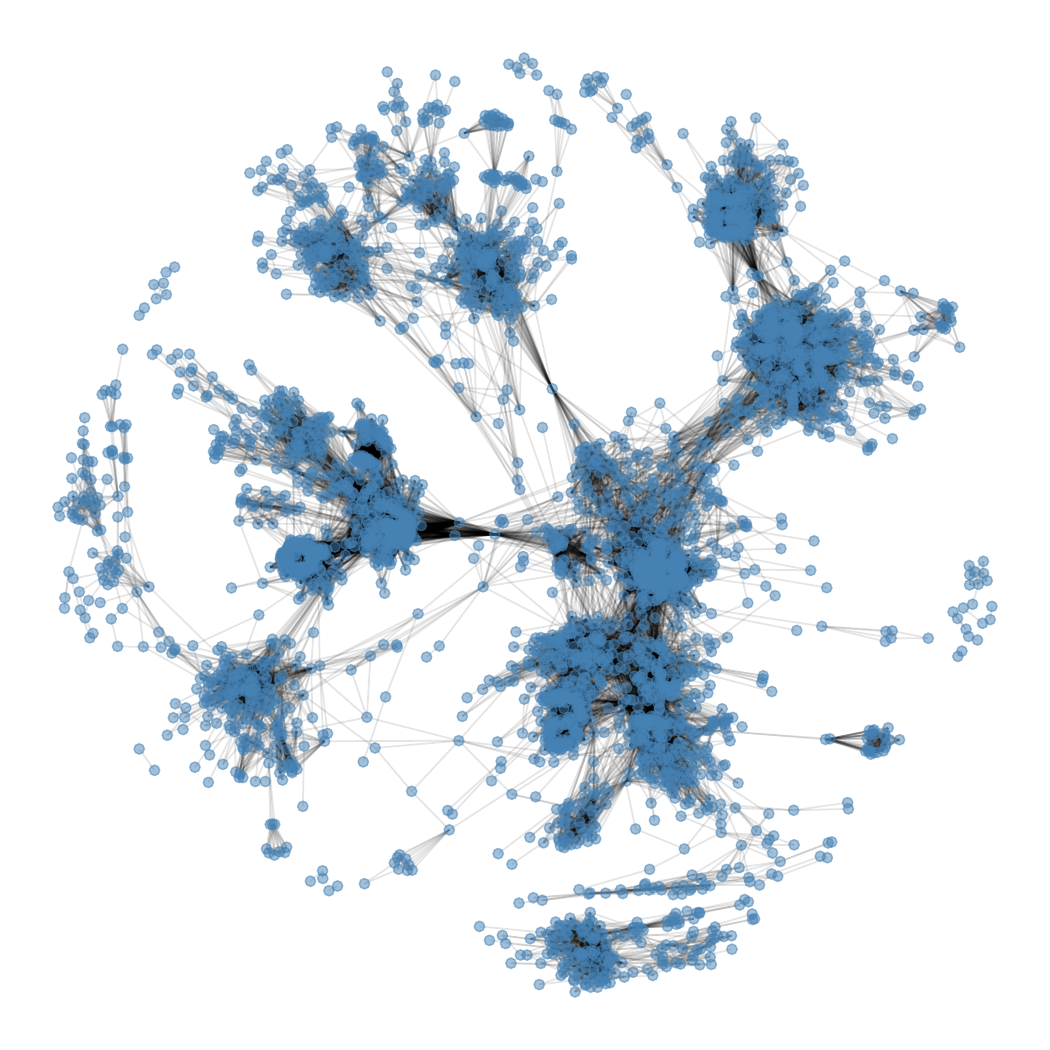}
\caption*{Illustration of the network. The left plot depicts the complete network with 4039 nodes including the egos. The right plot depicts the network with 3953 nodes after exclusion of the egos and all nodes only having connections to egos.}
\end{figure*}

Besides the challenges in estimation, a  critical shortcoming of the ERGM in its classical formulation \eqref{eq:ergm} is that it assumes node homogeneity, given the network statistics. In other words, the actors in the network can be permuted without changing the model.
In practice, this can be a crucial caveat, in particular when modelling large networks, and the inclusion of latent node heterogeneity in the model is advisable. 
Accounting for such heterogeneity traces back to the so-called $p_1$ and $p_2$ models, which explain the existence of an edge purely with external nodal covariates or random effects, respectively \citep{HolLei1981,DuiSniZij2004}. In this model setup, \citet{doi:10.1080/01621459.2023.2187815} recently proposed  variable selection in the presence of node heterogeneity. The ideas of node heterogeneity  were extended in \citet{Kos2009} and further on in \citet{ThiKau2016}, who combined the node-specific random effects of the $p_2$ model with ERGMs and proposed  Bayesian estimation via MCMC. While conceptually straight-forward, the MCMC approach becomes numerically infeasible in larger networks. 
As an illustration, take the large undirected social network depicted in Figure~\ref{fig:facebook} with about 4,000 nodes and nearly 90,000 edges, a set of Facebook profiles with edges representing  friendships. Heterogeneity in the nodes and their connectedness is directly visible. We will account for this by including latent node-specific heterogeneity.
 \citet{BoxChrMor2018} propose approximating estimation through a pseudolikelihood approach, which does work in large networks, but has its own problems such as  bad behaviour or void inference \citep[see e.g.,][]{DujgilHan2009}. \cite{KevKau2021} propose  pseudolikelihood  estimation in networks with nodal heterogeneity and extend it to bipartite networks in \cite{KevKau2022}. However, this suffers from biased inference. 

If node- or edge-specific covariates, denoted as $x$, are also available, including these in the model can  reduce nodal heterogeneity. In this case, \eqref{eq:ergm} extends to
\begin{align}
\label{eq:ergm2}
\dsP(Y=y\mid \beta) &= \frac{ \exp( s(y,x)^\top \beta) }{\kappa(\beta)},
\end{align}
where $s(y,x)$ is a vector of statistics built from both the endogenous $y$ and the  exogenous components $x$. The main assumption of model (\ref{eq:ergm2}) is that all nodal heterogeneity is captured through the exogenous covariates $x$. As such, $x$ can be seen as a set of confounding variables that capture node-specific heterogeneity. However, in real applications, one might not have included all covariates to explain heterogeneity; that is to say, some nodal heterogeneity may remain, which cannot be explained through the included confounders. We will therefore also  extend model \eqref{eq:ergm2} towards including additional nodal heterogeneity to account for such effects not explained by  $x$.  

If random nodal heterogeneity is required, covariates $x$ do not fully explain differences between the nodes. Vice versa, if the nodal heterogeneity is low or not required in the model with covariates, nodal heterogeneity is fully explained by the external covariates as generally desired for interpretability. Taking this view, we may state that reducing nodal heterogeneity through the inclusion of covariates $x$ allows quantification of the amount of {\sl posterior explained variance}, comparable to the coefficient of determination $R^2$ in classical regression models.  We facilitate this idea through Bayesian reasoning using versatile penalized complexity priors \citep{KleKne2016,SimRueMarRieSor2017} on variance parameters of the priors for model coefficients associated with the nodal random effects.

{We illustrate  this extended  approach in two further examples. Figure~\ref{fig:parliament} visualizes the considered  collaboration network of the members of the Italian Parliament in an undirected network with edge-specific covariates.  
 \begin{figure}[htbp]
\centering
\caption{Political collaboration network.}\label{fig:parliament}
\includegraphics[width=0.45\textwidth]{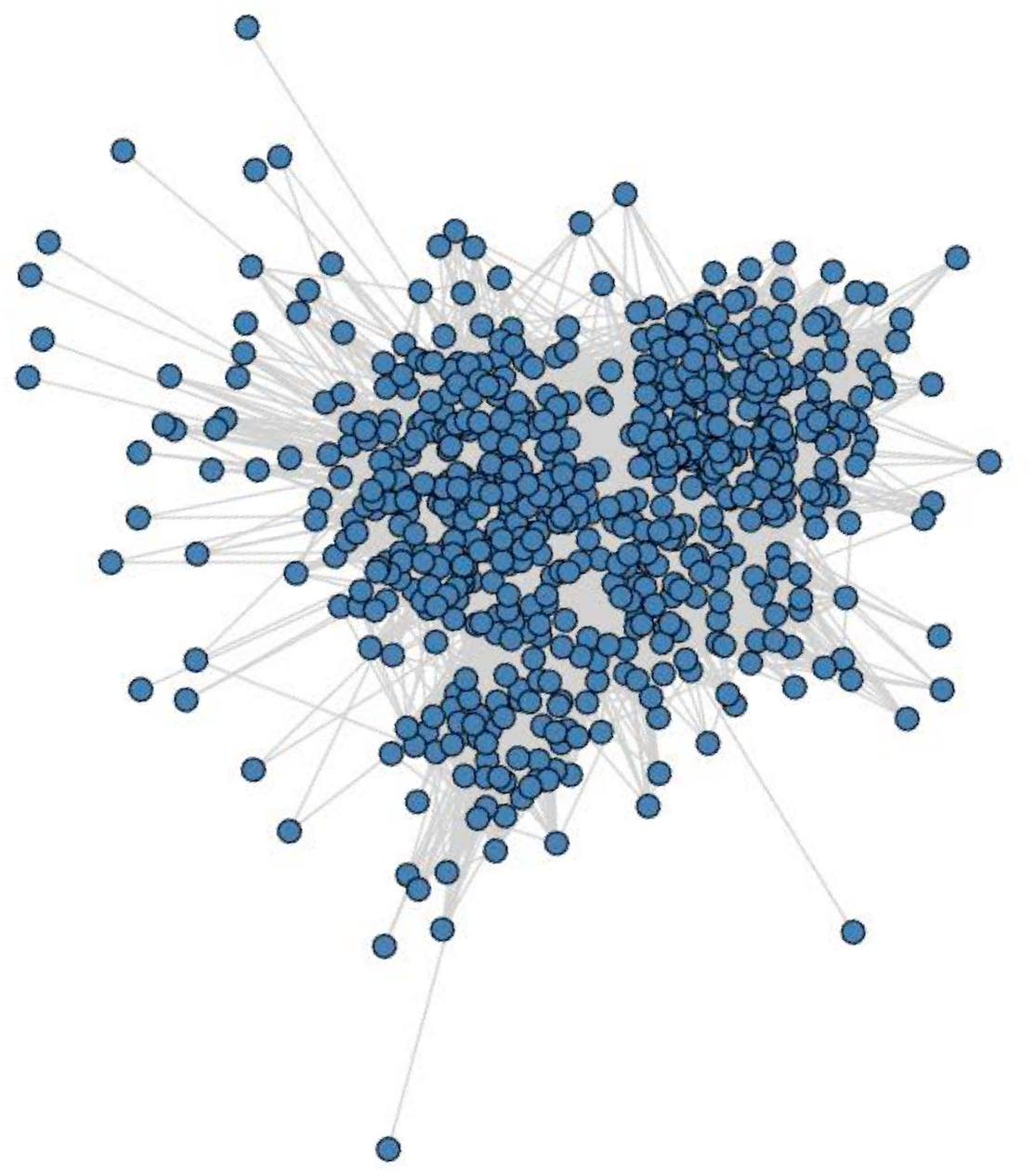}
\caption*{Illustration of the network. It represents the Italian Parliament using a  binary network with an edge representing at least one co-sponsorship.}
\end{figure}
Next, we consider is a directed network of international arms trading in 2016. Figure~\ref{fig:weapons} illustrates the respective network with directed edges indicating the export directions. 
\begin{figure*}[htbp]
\caption{International arms trading.}\label{fig:faebook}
\centering
\includegraphics[width=0.8\textwidth]{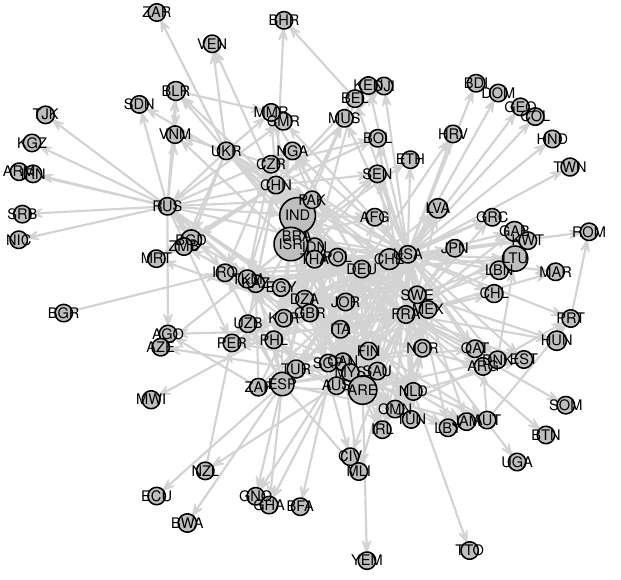}
\caption*{Illustration of the network. Each node represents a country, with all country abbreviations explained in Table~E.1 in the supplement. The size of the nodes relates to the outdegree of each node (the more outgoing edges, the larger the node). The arrows indicate the export directions.
}\label{fig:weapons}
\end{figure*}
We showcase that controlling for important confounders through the inclusion of node-specific covariates (such as gross domestic product) can reduce nodal random effects, thus providing a way of reducing unobserved heterogeneity. This observation is confirmed in a simulation study with nodal covariates and is not necessarily the case for edge-specific covariates, as we will see in the parliament data.}

To make estimation feasible, even  for  large networks with nodal heterogeneity, we propose the use of variational inference (VI) as a scalable and stable alternative to existing estimation procedures for ERGMs. 
We propose a computationally efficient  variational Bayes (VB) estimator~\citep{JorGhaJaaSau1999,OrmWan2010}
to compute approximate posterior inference quickly, even when the
network and parameter dimension are large. VI is an optimization-based technique for approximate Bayesian inference that has gained in popularity over the last few years, as it provides a computationally efficient alternative to sampling methods that scales well to high-dimensional problems and highly parameterized models \citep[][]{BleKucMcA2017}. Variational methods typically seek to minimize the Kullback–Leibler
(KL) divergence between the possibly complex true posterior and a more tractable so-called variational
density, which can be shown to be equivalent to maximizing a lower bound $\mathcal{L}$
on the log-marginal likelihood. 
We  consider stochastic VI \citep[SVI; cf.~e.g.~][]{NotTanVilKoh2012,PaiBleJor2012,SalKno2013}, which does not require
expectations to be evaluated analytically, but enables the use of an unbiased Monte Carlo estimate of the gradient of $\mathcal{L}$ to make inference tractable.
The VB estimator is based on a Gaussian approximation with a sparse factor representation of its covariance matrix~\citep{OngNotSmi2018} 
and applicable to 
ERGMs with node heterogeneity even for large networks. It has proven to be highly  accurate in a number of simulations conducted. These simulations also allow us to define suitable default values for involved tuning and optimization options. To further facilitate the use of our approach, we also propose feasible options to answer model-fit and model-selection questions. Variational approximations in ERGMS have been used before in 
 \citet{MelZhu2021}, who propose a mean-field-type variational approximation, and \citet{TanFri2020}, who develop a variety of variational methods for ERGMs, but their respective models do not account for heterogeneity. For other network models with block or mixture components, variational  methods have been used in e.g.,   \citet{LatBirAmb2012,McDBreFriHur2013}, or more recently in \citet{babkin2020} and \citet{Yin-etal:2022}.

Overall, we view the contributions of this paper as follows. First,  our VI approach scales up to high dimensions and highly-parameterized models, in which available estimation techniques for network data analysis meet their limits. We are thus able to estimate networks with thousands of nodes and accordingly thousands of nodal random heterogeneity effects. Secondly, if node-specific covariates  are available, we liaise posteriors of nodal random effects  for model validation and model selection, which indeed allows us to quantify the amount of (posterior) explained uncertainty, similar to classical linear  regression models. Finally, we also propose heuristic amendments for inference based on standard error corrections of our variational approximation, leading to more reliable quantification of uncertainty.  

 
  We begin with a review on directed and undirected Bayesian ERGMs with nodal random effects and make our choices for prior distributions in Section~\ref{sec:MERGM}. Section~\ref{sec:VB} details our estimation approach  in these models through VB, while Section~\ref{sec:modchoice} discusses practical options for model selection.  Three up-to-date illustrations on network data from the Italian Parliament, international arms trading, and Facebook egos in Section~\ref{sec:empirical} underpin the benefits of VB for ERGMs including heterogeneity.  Section~\ref{sec:conclusion} concludes. A detailed supplement contains the employed network statistics (A), further derivations and calculations relevant to our VI approach (B--D), additional material on the  real-data examples (E), and extensive simulations (F). To explore the performance of our VI approach, we specifically investigate its accuracy and robustness, validity of (corrected) uncertainty estimates, model selection through approximate Bayes factors and posterior explained variance, as well as benchmarking with exact Bayesian inference in small networks. Overall, the simulation results are satisfying.

\section{Bayesian  ERGMs with Nodal Random Effects}\label{sec:MERGM}

This section describes our Bayesian ERGM with nodal random effects.  We review  ERGMs with latent nodal heterogeneity in Section~\ref{sec:mergm} for undirected and directed graphs.  Our prior choices used for Bayesian inference are discussed in Section~\ref{sec:prior}. For notational simplicity we omit possible covariates $x$ in the notation and  write the statistics as $s(y)$ even though all results directly apply to statistics of the form $s(y,x)$ as in \eqref{eq:ergm2}.

\subsection{Mixed ERGMs}\label{sec:mergm}
To compensate for heterogeneity in the nodes of a network in ERGMs of the form \eqref{eq:ergm}, \cite{ThiKau2016} consider combining the $p_2$ model of~\citet{DuiSniZij2004} with \eqref{eq:ergm} to a so-called mixed ERGM of the form
\begin{equation}\begin{aligned}
\label{eq:mergm}
p(y\mid\beta,\gamma)&\coloneqq\dsP\left(Y=y\mid\beta,\gamma\right) \\&= \frac{ \exp( s(y)^\top \beta + t(y)^\top\gamma) }{\kappa(\beta,\gamma)},
\end{aligned}\end{equation}
where  $t(y)$ contains the degree statistics of the $N$ vertices. For an undirected graph, $t(y)=(t_1(y),\ldots,t_N(y))^\top$  counts the  ties for each node $i=1,\ldots,n$,  i.e~$t_i(y)=\sum_{j=1}^N Y_{ij}$. 
The associated coefficient vector $\gamma=(\gamma_1,\ldots,\gamma_N)^\top$ is assumed to be normally distributed, 
$\gamma_i\overset{\mbox{\scriptsize{i.i.d.}}}{\sim}N(\mu_\gamma,\sigma_{\gamma}^2)$, where $\ND(\mu,\sigma^2)$  denotes a Gaussian distribution with mean $\mu$ and variance $\sigma^2$. The parameter $\mu_\gamma$ thereby captures the average propensity in the network to form a tie. Therefore, $\beta_0$, which is usually the parameter associated with the edge statistic $\beta_{\mbox{\scriptsize{edge}}}\equiv\beta_0$, is excluded from $\beta$, i.e.~$\beta = (\beta_1,\ldots, \beta_p)^\top$. Alternatively, one can set $\mu_\gamma=0$. Indeed, there is the relation $\beta_{\mbox{\scriptsize{edge}}}=2\mu_\gamma$  \citep{ThiKau2016}. In the undirected case, the likelihood in \eqref{eq:mergm} arises from  the conditional formulation
\begin{equation}\begin{aligned}
    \label{eq:cond}
\mbox{logit}\left\lbrack \dsP\left(Y_{ij}=1 \mid Y_{kl},\, (k,l)\neq (i,j);\,\beta,\gamma_i,\gamma_j\right)\right\rbrack \\
= s_{ij}(y)^\top\beta+\gamma_i+\gamma_j,
\end{aligned}\end{equation}
for $1\le i \ne j \le N$ and 
where $s_{ij}(y)$ denotes the vector of so-called change statistics
\[
s_{ij}(y) = s(y_{ij} = 1\mid Y_{-ij})-s(Y_{ij} = 0 \mid Y_{-ij}),
\]
with $Y_{-ij}$ being the status of the remaining network except for  the edge between node $i$ and $j$.
 When the network is directed, the statistics in model \eqref{eq:mergm} extend to 
\begin{equation*}\begin{aligned}
 t(y)&=(t_{11}(y),\ldots,t_{1N}(y),t_{21}(y),\ldots,t_{2N}(y))^\top,\\\gamma&=(\delta_{1},\ldots\delta_{N},\phi_{1},\ldots,\phi_{N})^\top,
\end{aligned}\end{equation*}
where $t_{1i}(y)=\sum_{j=1}^N Y_{ij}$, $t_{2i}=\sum_{j=1}^N Y_{ji}$ count the incoming and outgoing edges connected to node $i$, while $\delta_i$ and $\phi_i$ are the corresponding sender and receiver effects, respectively. Similarly to the undirected case, $(\delta_i, \phi_i)^\top$ are random  and \eqref{eq:cond} takes the form
\begin{equation*}\begin{aligned}
\mbox{logit}\left\lbrack \dsP\left(Y_{ij}=1 \mid Y_{kl},\, (k,l)\neq (i,j);\,\beta,\delta_i, \delta_j,\phi_i,\phi_j\right)\right\rbrack \\= s_{ij}(y)^\top\beta+\delta_i
+ \phi_j 
\end{aligned}\end{equation*}
where additional constraints on the mean values of $\delta_i$ and $\phi_j$ are required, as detailed below.

\subsection{Prior distributions}\label{sec:prior}

We extend the above models now towards a  Bayesian treatment by formulating appropriate prior distributions. For the coefficient vector $\beta$ of network statistics and the vector of nodal random effects $\gamma$, we employ Gaussian priors with appropriate covariance structures, depending on whether the network is directed or undirected.
We then make further prior assumptions on the covariances. 

\begin{figure*}[t]
    \centering
    \includegraphics[width=0.8\textwidth]{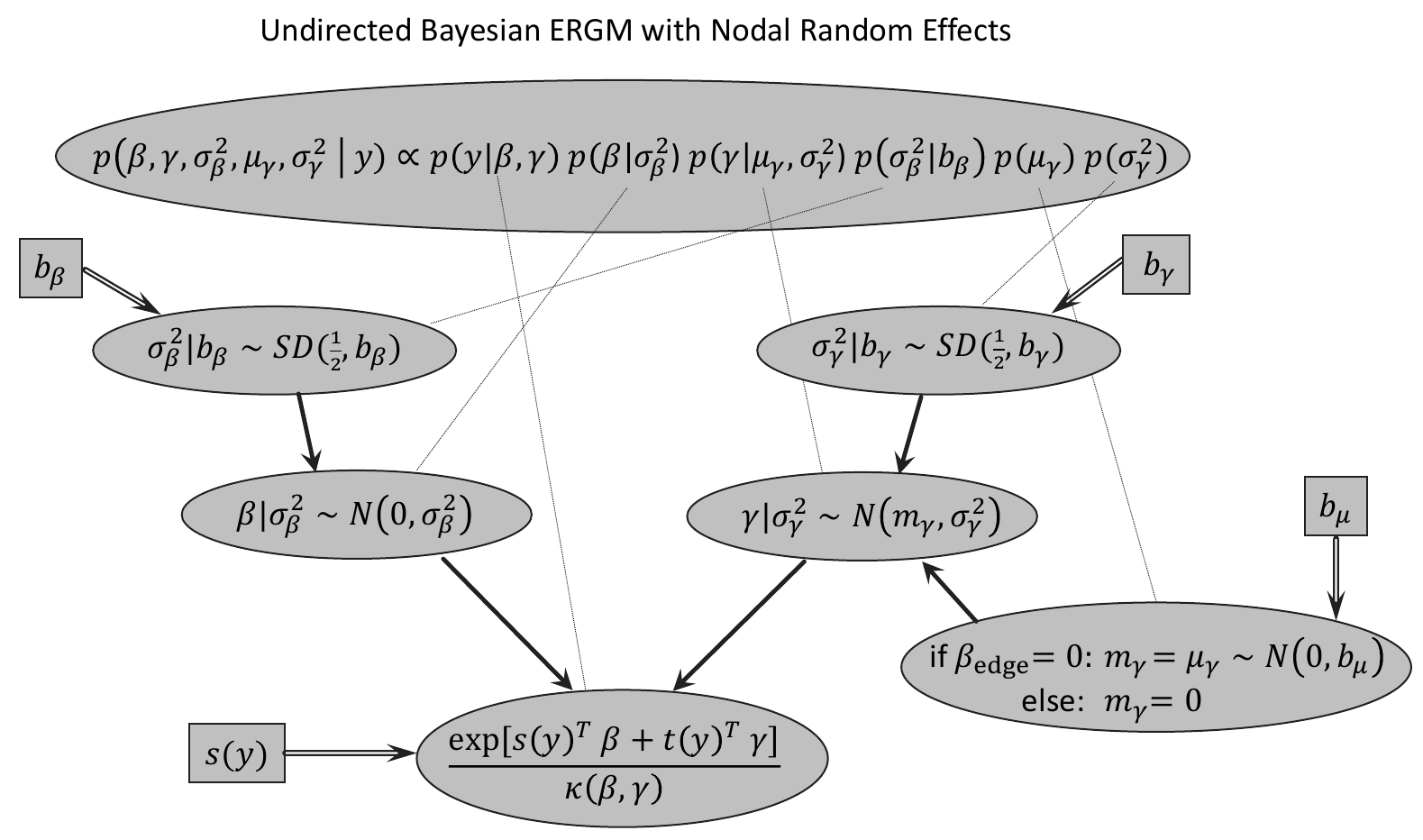}
    \caption{\footnotesize Graphical representation of the undirected Bayesian ERGM with nodal effects. Ellipses are stochastic nodes and rectangles are deterministic/logical
nodes. Single arrows are stochastic edges and double arrows are logical/deterministic edges. In simulations in Section F, the cases $\beta_{\mbox{\scriptsize{edge}}}=0$ and $\mu_\gamma=0$ are compared and denoted with the suffix ``-het($\beta_{\mbox{\scriptsize{edge}}}=0$)'' and ``-het($\mu_\gamma=0$)'', respectively.}
    \label{fig:undir}
\end{figure*}

\paragraph*{Undirected Bayesian mixed ERGM}
In the undirected case, we have $\beta=(\beta_1,\ldots,\beta_p)^\top$ and $\gamma=(\gamma_1,\ldots,\gamma_N)^\top$. As mentioned before, it is commonly assumed  that $\beta_0\equiv 0$, such that the intercept (edge statistic) is implicitly  modelled through the mean $\mu_\gamma$ of the Gaussian prior for $\gamma$. Overall, we make the following prior assumptions:
\begin{equation*}\begin{aligned}
\beta&\sim \ND(0,\sigma_\beta^2 I_{p\times p}), \mbox{  } &\sigma_\beta^2&\sim \mbox{SD}(1/2,b_\beta)\\
\gamma&\sim \ND(\mu_\gamma,\sigma_\gamma^2 I_{N\times N})
, \mbox{ }
\mu_\gamma\sim \ND(0,b_\mu), \mbox{  } &\sigma_\gamma^2&\sim \SDD(1/2,b_\gamma),
\end{aligned}\end{equation*}
where  $
\mbox{SD}(1/2,b)$ denotes the scale-dependent prior of~\citet{KleKne2016} (which corresponds to a Weibull prior with shape parameter $1/2$ and scale parameter $b$), respectively. {This  prior relies on the principles of penalized complexity priors \citep{SimRueMarRieSor2017} and has the appealing property of penalizing from a flexible alternative to a properly defined base model. For our case, this corresponds to the situation in which evidence in the data should decide between a nested model without nodal random effects (base model) and a more flexible alternative containing nodal random effects, a topic we further investigate in the presence of nodal covariates in Section~\ref{subsec:weapons}. Empirically, our simulations are rather robust against the specific choices of hyperparameters $b_\mu$, $b_\beta$, and $b_\gamma$, such that we set $b_\mu=100$, $b_\beta=100$ and $b_\gamma=100$, 
 see also Supplement~F.1 for empirical evidence on robustness against these choices.} We denote the vector of all model parameters in the undirected Bayesian mixed ERGM with dimension $\dim(\vartheta)=p_\vartheta=p+N+3$ as
\begin{equation}\label{eq:theta1}
\vartheta=( \beta^\top,\gamma^\top,\sigma_\beta^2,\mu_\gamma,\sigma_\gamma^2)^\top.
\end{equation}
Figure~\ref{fig:undir} presents a graphical illustration of the undirected Bayesian mixed ERGM.

\paragraph*{Directed Bayesian mixed ERGM}
In the directed case, we have $\beta=(\beta_0,\beta_1,\ldots,\beta_p)^\top$ and $\gamma=(\delta_1,\ldots,\delta_N,\phi_1,\ldots,\phi_N)^\top$. In this setting, it is notationally more convenient to include $\beta_0$ as  edge effect. 
While the priors for $\beta$ and $\sigma_\beta^2$  are the same as in the undirected case, the hierarchical prior for $\gamma$ is
\begin{equation*}\begin{aligned}
\gamma&\sim \ND(0,\Sigma), \\
\Sigma&=\underbrace{\begin{bmatrix}  \sigma_{\delta}^2 &  \sigma_{\delta\phi}^2\\ \sigma_{\delta\phi}^2  &\sigma_{\phi}^2 \end{bmatrix}}_{\Sigma_{\delta,\phi}} \otimes I_N =\begin{bmatrix} \sigma_\delta^2 & \sigma_\delta\sigma_\phi\rho \\ \sigma_\delta\sigma_\phi\rho & \sigma_\phi^2 \end{bmatrix} \otimes I_N, \\
\sigma_\delta^2&\sim{\SDD}(1/2,b_\delta),\mbox{ }
\sigma_\phi^2\sim\SDD(1/2,b_\phi), \mbox{ } \rho\sim \mbox{Beta}(a_\rho,b_\rho),
\end{aligned}\end{equation*}
where $\mbox{Beta}(a,b)$ denotes a beta distribution with parameters $a,b$.
In all our simulations and applications, we set {$b_\delta=b_\phi=100$}, $a_\rho=b_\rho=1$, resulting in weakly informative priors for the marginal variances and uniform priors for $\rho$. In similarity to the undirected case, posterior results are rather robust against exact choices for these hyperparameters (cf.~Supplement~F.1).
 Finally, the vector of all model parameters in the directed Bayesian mixed ERGM with dimension $p_\vartheta=p+2N+5$ is
\begin{equation}\label{eq:theta2}
\vartheta=( \beta^\top,\gamma^\top,\sigma_\beta^2,\sigma_\delta^2,\sigma_\phi^2,\rho)^\top.
\end{equation}
For posterior estimation, some parameters are re-parameterized to the real line $\dsR$. Specifically, $\sigma_\beta^2$ is log-transformed, similar to $\sigma_\gamma^2$ (undirected case) and $\sigma_\delta^2,\sigma_\phi^2$ (directed case). For the directed model, a Fisher-$z$ transformation is  used for $\rho$, see  Supplement~B for resulting priors on the transformed parameters.

\section{Variational Inference for  Bayesian Mixed ERGMs}\label{sec:VB}

In this section, we review basic ideas of VI and introduce a scalable VI method for the Bayesian mixed ERGM. 

\subsection{Variational inference in a nutshell}
VI methods are a promising approach
to scaling approximate Bayesian inference to work in large
datasets and highly parameterized models. 
The general idea of VI is to approximate a posterior density $p(\vartheta|y)\propto p(y|\vartheta)p(\vartheta)=h(\vartheta)$  by a  more tractable density $q_\lambda(\vartheta)$, the so-called variational density. Here, $p(y|\vartheta)$ is the conditional
likelihood at \eqref{eq:mergm}, and $\lambda$ is a vector of 
{variational parameters} that are calibrated
by minimizing some measure of closeness between $q_{\lambda}(\vartheta)$ and $p(\vartheta|y)$.  Commonly, the measure of closeness adopted is the Kullback-Leibler divergence (KL) from $q_{\lambda}(\vartheta)$ to $p(\vartheta|y)$ given by $\mbox{KL}(\qlt\Vert p(\vartheta|y))=\int\log(\qlt/p(\vartheta|y))\qlt d\vartheta$. Then, writing $p(y)=\linebreak\int p(\vartheta)p(y|\vartheta) d\vartheta$ for the marginal likelihood, the following identity holds for any $q_\lambda(\vartheta)$:
\begin{equation}\begin{aligned}\label{eq:marg}
\log(p(y))&= \int\log\left(\frac{p(y,\vartheta)}{\qlt}\right)\qlt d\vartheta\\&\qquad +\mbox{KL}(\qlt\Vert p(\vartheta|y)).
\end{aligned}\end{equation}
Since the KL is non-negative,
\begin{equation}\begin{aligned}\label{eq:Elb}
\mathcal{L}(\lambda)&=\int q_{\lambda}(\vartheta)\log\left\lbrace\frac{p(y,\vartheta)}{q_{\lambda}(\vartheta)}\right\rbrace d\vartheta\\&=\int q_{\lambda}(\vartheta)\log\left\lbrace\frac{p(y|\vartheta)p(\vartheta)}{q_{\lambda}(\vartheta)}\right\rbrace d\vartheta\,,
\end{aligned}\end{equation}
is a lower bound on $\log(p(y))$ known as the variational lower bound or evidence lower bound (ELBO). Because the left-hand side of~\eqref{eq:marg} does not depend on $\lambda$, minimizing the KL with respect to $\lambda$ is equivalent to maximizing  $\mathcal{L}$ with respect to $\lambda$, which is a more convenient objective not involving the intractable marginal likelihood. Introductory overviews on variational inference can be found in \citet[][]{OrmWan2010,BleKucMcA2017}. 

\subsection{Stochastic variational inference}

Maximizing $\mathcal{L}({\lambda})$ directly  to  obtain  an optimal variational approximation (VA)  is  often  difficult, since $\mathcal{L}({\lambda})$ is defined as an integral that is generally intractable, as it is in our case.  However,~\eqref{eq:Elb} is  an expectation with respect to $q_\lambda(\vartheta)$ 
\begin{equation}\label{eq:Elb2}
\mathcal{L}(\lambda)=\dsE_{q_\lambda}\left\lbrace\log(h(\vartheta))-\log(q_{\lambda}(\vartheta))\right\rbrace.
\end{equation}
The expectation  \eqref{eq:Elb2} enables  unbiased Monte Carlo (MC) estimation, which extends to the estimation  of the gradient of $\mathcal{L}(\lambda)$ after differentiating under
the integral sign.  Doing so, the gradient $\nabla_\lambda\mathcal{L}(\lambda)$ results  in an expectation with respect to $q_{\lambda}$,
\begin{equation}\label{eq:dElb2}
\nabla_\lambda\mathcal{L}(\lambda)=\dsE_{q_\lambda}\left\lbrace\nabla_\lambda\log(\qlt)[\log(h(\vartheta))-\log(q_{\lambda}(\vartheta))]\right\rbrace\,,
\end{equation}
where the so-called log-derivative trick $\dsE_{q_\lambda}(\nabla_\lambda\log(\qlt))=0$ was employed.
This is often used with stochastic gradient ascent (SGA) methods and there is now a large literature surrounding the application of  this  idea, also called stochastic variational inference~\citep[SVI; see][among others]{NotTanVilKoh2012,PaiBleJor2012,HofBleWanPai2013,SalKno2013,KinWel2014,RezMohWie2014,RanGerBle2014,TitLaz2015,KucTraRanGelBle2017}.  Denoting with $\reallywidehat{\nabla_{\lambda}\mathcal{L}(\lambda)}$  an unbiased MC estimate of the gradient $\nabla_\lambda\mathcal{L}(\lambda)$, SGA methods proceed as follows. Choose an initial value $\lambda^{(0)}$ and for $t=0,1,\ldots$ perform the update
\begin{equation}\label{eq:SGA}
\lambda^{(t+1)}=\lambda^{(t)} + \nu^{(t)}\circ\reallywidehat{\nabla_{\lambda}\mathcal{L}(\lambda^{(t)})}\,
\end{equation}
recursively, where $\circ$ denotes the Hadamard (element-by-element) product
of two random vectors and $\lbrace \nu^{(t)} \rbrace_{t\geq 0}$ 
is a sequence of vector-valued learning rates with dimension $\dim(\lambda)$. The update is continued until a stopping
condition is satisfied. The learning rate sequence is  chosen to satisfy the Robbins–Monro conditions $\sum_t\nu_j^{(t)}=\infty$ and $\sum_t(\nu_j^{(t)})^2<\infty$~\citep{RobMon1951},  $j=1,\ldots,\dim(\lambda)$, which ensures that the iterates $\lambda^{(t)}$ converge to a (possibly local) optimum as $t\xrightarrow{}\infty$ under suitable regularity conditions~\citep{Bot2010}. In practice, it is important to consider
adaptive learning rates, and these are set here  
 using the ADADELTA method of~\cite{Zei2012}, as in~\cite{OngNotSmi2018}.

For the SGA algorithm to be efficient, the estimate $\reallywidehat{\nabla_{\lambda}\mathcal{L}(\lambda)}$ 
should exhibit low variance, and the references above on SVI differ in the way an unbiased estimate $\reallywidehat{\nabla_{\lambda}\mathcal{L}(\lambda)}$  is constructed and how its variance is reduced. Here we consider  the so-called  re-parameterization trick \citep{KinWel2014,RezMohWie2014}. \citet{XuQuiKohSis2019} recently provided evidence that employing this method is  often much more efficient than existing alternatives. We give details on  how to employ the re-parameterization trick in our case in  Supplement~C.2.

\subsection{Choice of the variational approximation}
Successful application of the variational methods described above requires a suitable and numerically tractable approximation density $q_{\lambda}(\vartheta)$. 
We follow 
\cite{OngNotSmi2018} and employ a Gaussian 
 variational approximation $q_{\lambda}(\vartheta)=\phi(\vartheta;\mu,\Upsilon)$, where the covariance matrix is a parsimoniously structured factor covariance matrix with $K$ factors. This choice allows for a flexible full covariance structure, while keeping the number of unknowns feasible. Furthermore, the Gaussian  approximation enables the use of the aforementioned re-parameterization trick. We refer to Supplement~C.1--C.3 for details on the variational density, variance reduction through the re-parameterization trick, as well as implementation of the SGA and involved gradients.  From Supplement~F.1, we choose $K=20$ as our default, yielding a good balance between accuracy and computational efficiency.

\subsection{Estimation of $E_{\beta,\gamma}(s({y}))$ and $E_{\beta,\gamma} (t({y}))$}
Computation of the gradients with respect to $\beta$ and $\gamma$ involves the derivatives $\frac{\partial \log(\kappa(\beta,\gamma))}{\partial \beta}$ and $\frac{\partial \log(\kappa(\beta,\gamma))}{\partial \gamma}$, which are --- similar to $\kappa(\beta,\gamma)$ --- themselves analytically intractable, since 
\begin{align*}
\frac{\partial \log(\kappa(\beta,\gamma))}{\partial \beta} 
& = \underbrace{\frac{\sum_{y \in \mathcal{Y}}\exp(\beta^\top s(y)+\gamma^\top t(y))}{\kappa(\beta,\gamma)}}_{p(y|\beta,\gamma)}s(y)\\&=\sum_{y \in \mathcal{Y}}p(y|\beta,\gamma)s(y)=\mathbb{E}_{y|\beta,\gamma}[s(y)]\\
\frac{\partial \log(\kappa(\beta,\gamma))}{\partial \gamma} 
&= \underbrace{\frac{\sum_{y \in Y}\exp(\beta^\top s(y)+\gamma^\top t(y))}{\kappa(\beta,\gamma)}}_{p(y|\beta,\gamma)}t(y)\\
&=\sum_{y \in Y}p(y|\beta,\gamma)t(y)=\mathbb{E}_{y|\beta,\gamma}[t(y)].
\end{align*}
\citet{TanFri2020} consider three options for approximating $\mathbb{E}_{y|\beta,\gamma}[s(y)]$ and $\mathbb{E}_{y|\beta,\gamma}[t(y)]$ numerically, namely MC sampling and (adaptive) importance sampling ((A)IS). In our simulations (F.1), we find that (A)IS is the slightly faster method for very small parameter spaces. However, the larger the parameter space, the worse the approximation quality of (A)IS becomes. Even  for a network of only 40 nodes with random effects, MC sampling is more efficient. These findings are in line with the statements made in \citet{TanFri2020}; we therefore employ MC sampling as default. 
At each iteration, $L$ networks are simulated using the ``tie-no-tie'' sampler in the \texttt{ergm} package \citep{HunGooHan2008}, which is a Metropolis-Hastings sampler. As the estimation only needs to be unbiased, it is sufficient to sample  $L=5$ networks (with burn-in phase of 5000), as demonstrated empirically in  Supplement~F.1.

\subsection{Uncertainty quantification}\label{sec:cred}
In principle, standard errors for $\hat\vartheta$ can be directly computed using the VA $q_{\hat\lambda}=\ND(\hat\mu,\hat\Upsilon)$. Even though it has been demonstrated  in some situations that VI does not suffer from accuracy in terms of posterior predictive densities \citep[e.g.,~][]{BraMcA2010,KucTraRanGelBle2017}, it is known that VI generally underestimates the variance of the posterior density \citep{BleKucMcA2017}, which is a
consequence of its objective function. 
A few papers try to match more closely the inferences made by MCMC \citep[e.g.,~][]{GioBroJor2018}. However, these are often computationally expensive or preliminary designed for mean field VB, which is different from our stochastic fixed-form VI \citep[compare][for a general discussion]{miller2021}.

To improve the coverage of credible intervals of $\theta=(\beta^\top,\gamma^\top)^\top\in\dsR^{p_{\theta}}$ derived from our $q_{\hat\lambda}$, we employ the following simple covariance correction for $\hat\Upsilon$.
We define with 
$
V:=
F_\theta^{-1}=\left[\Var\left(s(y), t(y)\right)^\top\right]^{-1}
$
the inverse Fisher information, which is estimated through 
\begin{equation}\label{eq:Fisher}
\hat V\:= \left(\frac{1}{B}\sum_{b=1}^B(s(y^b),t(y^b))^\top(s(y^b),t(y^b))\right)^{-1}
\end{equation}
based on $B$ simulated networks $y^1,\ldots,y^B$ from $p(y\:| \:\hat{\beta},\hat\gamma)$. We further define  $S=\diag(s_1,\ldots,s_{p_{\theta}})$ as the diagonal matrix, such that $s_i=\hat{V}_{ii}/\hat\Upsilon_{ii}$ holds for $i=1,\ldots p_\vartheta$. We now construct a corrected version of the full covariance matrix $\hat\Upsilon$ as
$
\hat\Upsilon_{\mbox{\scriptsize{corrected}}}=S^{1/2}\hat\Upsilon S^{1/2}.
$
We  investigate the coverage derived from $\hat\Upsilon$ and $\hat\Upsilon_{\mbox{\scriptsize{corrected}}}$ in Supplement~F.3. {The main conclusion is that applying the suggested correction generally leads to coverages closer to the nominal levels, with the tendency that $\hat\Upsilon$ underestimates the standard errors, while $\hat\Upsilon_{\mbox{\scriptsize{corrected}}}$ yields slightly too conservative credible intervals.}

An alternative way to construct credible intervals for $\theta$ would be to 
re-estimate the  networks $y^1,\ldots,y^B$ to obtain a sample $\lbrace\hat\theta^1,\ldots,\hat\theta^B\rbrace$ from which credible intervals could be constructed. However, if the networks are large, this can also be computationally demanding, since it requires the training of $B$ additional networks.

\subsection{VI algorithm  for Bayesian mixed ERGMs}
Algorithm~C.2 in Supplement~C.6 calibrates our proposed VA to the augmented posterior using SGA with the re-parameterization trick and the ADADELTA learning rate. The choice of starting values for $\vartheta$, stopping criterion, and computation of point estimates is detailed in Supplement~C.4 and C.5. In  Supplement~F.2, we  evaluate our VI approach to mixed ERGMs and compare it to exact inference via MCMC as proposed in \citep{ThiFriCaiKau2016} using  \texttt{mixbergm} \citep{mixbergm} and which is computationally feasible for small networks only. We find that here VI performs en par with MCMC, while being considerably faster and feasible when MCMC is not.

\section{{Bayesian Inference, Model Selection \& Goodness-of-Fit}}\label{sec:modchoice}

Important yet challenging aspects in network analyses are to find the best specification of the model and to evaluate the overall model fit. For mixed ERGMs, it is not only of interest to decide which network statistics to include but also to decide whether nodal random effects should be present in the model or not. In other words, we question whether the statistics included fully describe the network structure or whether unobserved nodal heterogeneity remains. 
To tackle these  questions, we propose three approaches. Firstly, we use the marginal posteriors of variances of the random nodal effects as a graphical device to measure the node heterogeneity across competing models. Secondly, we  rely on (approximate) Bayes factors \citep[BFs;][]{KasWas1995}. Thirdly, we propose  Bayesian goodness-of-fit \citep[BGOF;][]{HunGooHan2008} measures.

\subsection{\mbox{Explained variance through exogenous covariates}}

 The marginal posteriors of variances associated with the nodal random effects (i.e.~$\sigma_\gamma^2$ in the undirected cases  and $\sigma_{\delta}^2,\sigma^2_{\phi}$ in the directed cases, respectively) allow us to quantify the unexplained variability, i.e.,~the heterogeneity that cannot be explained by the network statistics $s(y)$. This becomes particularly interesting in the presence of node and/or edge-specific covariates $x$, as in model \eqref{eq:ergm2}. We may then investigate the influence of including the covariates on the heterogeneity of the nodes. We propose to do this by regarding  the differences in the respective posteriors in models fitted with and without covariates (i.e.,~$s(y,x)$ vs $s(y)$). This exhibits how much variability can be explained (or vice versa is induced) by including $x$ and defines a novel (graphical) measure to quantify the extent of nodal heterogeneity in competing network models in the presence of covariates that has not been used before. To be specific, we define the \textit{posterior explained variance} as follows. 
\begin{definition}[Posterior explained variance]\label{def1}
Let $\bullet\in\lbrace \gamma,\delta,\phi\rbrace$ and $\mathcal{M}_1^\ast$, $\mathcal{M}_2^\ast$ be two models with nodal random effects and the same network statistics but  without and with  exogenous  covariates $x$, respectively. Let $p_{\sigma_{\bullet,\mathcal{M}_1^\ast}^2\mid y}$, $p_{\sigma_{\bullet,\mathcal{M}_2^\ast}^2\mid y}$ be the respective marginal posteriors. Then, we define the \textit{posterior explained variance as}
\begin{equation}\begin{aligned}\label{eq:postvarexp}
    R_{\bullet}&=\dsP(\sigma_{\bullet,\mathcal{M}_2^\ast}^2<\sigma_{\bullet,\mathcal{M}_1^\ast}^2\mid y)\\&= \int_0^{\infty} \left\{ \int_{0}^s p_{\sigma_{\bullet,\mathcal{M}_2^\ast}^2\mid y}(t) dt \right\}\; p_{\sigma_{\bullet,\mathcal{M}_1^\ast}^2\mid y}(s) ds\\
   & \approx \int_0^{\infty} \left\{ \int_{0}^s q_{\sigma_{\bullet,\mathcal{M}_2^\ast}^2}(t) dt \right\}\; q_{\sigma_{\bullet,\mathcal{M}_1^\ast}^2}(s) ds,
   \end{aligned} \end{equation}
where $q_{\sigma_{\bullet,\mathcal{M}_1^\ast}^2}$, $q_{\sigma_{\bullet,\mathcal{M}_2^\ast}^2}$ are the respective marginal variational densities.
\end{definition}
The applicability of the posterior explained variance is investigated empirically in Supplement F.4, where we demonstrate the ability to reduce heterogeneity when the model contains nodal covariates.  We also exemplify this in Section~\ref{subsec:weapons}. However, when covariates are edge-specific, a reduction of the heterogeneity is not guaranteed, as we will see in Section~\ref{sec:parliament}.

\subsection{Approximate Bayes factors}\label{sec:BF}
The BF for comparing any two models $\mathcal{M}_1$ and $\mathcal{M}_2$, is defined as 
\[
\mbox{BF}_{12} = \frac{p(y|\mathcal{M}_1)}{p(y|\mathcal{M}_2)}, 
\]
where $p(y|\mathcal{M}_j)=\int p(y|\vartheta,\mathcal{M}_j)p(\vartheta|\mathcal{M}_j)\mathrm{d}\vartheta$, $j=1,2$ is  the marginal likelihood under $\mathcal{M}_j$ and   $p(y|\vartheta,\mathcal{M}_j)$ , $p(\vartheta|\mathcal{M}_j)$  are the respective likelihood and prior distributions. 
Generally, when $\mbox{BF}_{12} > 1$, the data favours $\mathcal{M}_1$
over $\mathcal{M}_2$. If $0<\mbox{BF}_{12}<1$, $\mathcal{M}_2$ is favoured instead. 

According to Bayes' theorem, the integration can be avoided and the  marginal likelihood $p(y)$ (suppressing the dependence on the model $\mathcal{M}$) can be obtained via
\begin{equation*}
\begin{aligned}
p(y) &= \frac{p(y,\vartheta)}{p(\vartheta|y)}=\frac{p(y|\vartheta)p(\vartheta)}{p(\vartheta|y)}
\end{aligned}
\end{equation*}
which involves the posterior $p(\vartheta|y)$ in the denominator. Let now  $\vartheta_1$, $\vartheta_2$ be the set of all model parameters under $\mathcal{M}_1$ and $\mathcal{M}_2$ with corresponding network statistics $s_1(\cdot)$ and $s_2(\cdot)$, respectively. Then, the BF can also be written as 
\begin{equation*}
    \begin{aligned}
    \mbox{BF}_{12}=\frac{\kappa(\beta_2,\gamma_2)}{\kappa(\beta_1,\gamma_1)}\frac{\exp(\beta_1^\top s_1(y)+\gamma_1^\top t(y))}{\exp({{\beta_2}}^\top s_2(y)+{\gamma_2}^\top t(y)) }
    \frac{p(\vartheta_1)}{p(\vartheta_2)}
    \frac{p(\vartheta_2|y)}{p(\vartheta_1|y)}.
    \end{aligned}
\end{equation*}
The posteriors  $p(\vartheta_1|y)$, $p(\vartheta_2|y)$ under $\mathcal{M}_1$, $\mathcal{M}_2$ are approximated by $q_{\hat\lambda_1}$ and $q_{\hat\lambda_2}$ evaluated at their point estimates $\hat\vartheta_1$ and $\hat\vartheta_2$, respectively, which we obtain from the variational approximations $q_{\hat\lambda_j}$ (see Supplement~C.5 for details). Plugging these also into the remaining terms, we  obtain our approximate BF denoted by $\reallywidehat{\mbox{BF}}_{12}$ as
\begin{equation}\label{eq:BF}
    \begin{aligned}
    \reallywidehat{\mbox{BF}}_{12}=\frac{\kappa({\hat\beta}_2,{\hat\gamma}_2)}{\kappa(\hat{\beta}_1,\hat\gamma_1)}
    \frac{\exp(\hat{\beta}^\top_1 s_1(y)+\hat{\gamma}^\top_1 t(y))}{\exp({\hat{\beta}}^\top_2 s_2(y)+{\hat{\gamma}}^\top_2 t(y)) }
    \frac{p(\hat\vartheta_1)}{p(\hat\vartheta_2)}
    \frac{q_{\hat\lambda_2}(\hat\vartheta_2)}{q_{\hat\lambda_1}(\hat\vartheta_1)}.
    \end{aligned}
\end{equation}
In \eqref{eq:BF}, all but the first factor ${\kappa({\hat\beta}_2,{\hat\gamma}_2)}/{\kappa(\hat\beta_1,\hat\gamma_1)}$ can be directly evaluated analytically. However, as the ratio of the normalizing constants is intractable, it needs to be approximated.
To do so, we use thermodynamic integration \citep[path sampling ;][]{CaiFri2013,ThiFriCaiKau2016}, and distinguish two situations:

\paragraph*{(I) Nested models}
If the models $\mathcal{M}_1$ and $\mathcal{M}_2$ are nested, two situations are possible, where w.l.o.g.~we assume that $\mathcal{M}_1$ is nested in $\mathcal{M}_2$.
\begin{itemize}
    \item[(I).1] $\mathcal{M}_1$ and $\mathcal{M}_2$ contain the same network statistics but $\mathcal{M}_2$ also accounts for nodal heterogeneity.  Assuming  $s_1(\cdot)$ and $s_2(\cdot) $  contain an intercept, we have $\gamma_1 \equiv 0$. 
    \item[(I).2] Both models $\mathcal{M}_1$ and $\mathcal{M}_2$ contain the sociality effects, 
    but $\mathcal{M}_2$ includes not only the network statistics of $\mathcal{M}_1$ but also further ones, i.e.~$s_1(\cdot)\subset s_2(\cdot)$ and $\dim(\beta_1)<\dim(\beta_2)$.
\end{itemize}
The path sampling approach of \citet{GelMen1998} can be applied  to  approximate numerically the expectation involved in computing $\log\lbrace{\kappa({\hat\beta}_2,{\hat\gamma}_2)}/{\kappa(\hat\beta_1,\hat\gamma_1)}\rbrace$  using the trapezoid rule  and repeated sampling from model \eqref{eq:ergm} on a grid in (0,1)  \citep[see Section~3.2 of][for details]{ThiFriCaiKau2016}.

\paragraph*{(II) Non-nested models}
If  models $\mathcal{M}_1$ and $\mathcal{M}_2$ are non-nested, augmenting $\beta_1$, $\gamma_1$ into  $\beta_2$, $\gamma_2$ 
is no longer possible, such that the usual path sampling approach cannot be applied. Instead, path sampling has to be performed twice, for $\log\lbrace\kappa(\vartheta_2)/\kappa(0)\rbrace$ and $\log\lbrace\kappa(0)/\kappa(\vartheta_1)\rbrace$ separately, with $\kappa(0)$ being the normalizing constant of a null model, in which all parameters are set to 0 \citep[see Section~3.3 of][for further details]{ThiFriCaiKau2016}. As computing time is almost doubled, (I) should be used whenever two models are nested. 

A generic representation of the computation of our BFs is provided in Algorithm D.1 in Supplement~D, while Section~F.5 in the Supplement~demonstrates that, overall, the BFs are able to decide correctly whether or not node heterogeneity is present and which network statistics should be retained in a model.

\subsection{Bayesian goodness-of-fit diagnostics}\label{sec:bgof}
\citet{HunGooHan2008}  consider the conditional distribution of ``out-of-model'' statistics at the MLE, and compare it to the observed graph.  Subsequent Bayesian work has done the same  with posterior predictive distributions as  adequacy checking. The latter is also referred to as Bayesian goodness-of-fit 
diagnostics (BGOFs) for ERGMs by comparing high-level statistics of observed networks with those of
corresponding networks simulated from the estimated network, in order
to evaluate the model fit in terms of posterior predictive assessment. The estimated posterior distribution of the model parameters is approximated by $q_{\hat\lambda}$ in our VI approach.  The set of statistics used for the comparison is the
\begin{itemize}
\item \emph{degree distribution}, which counts the nodes with degree $k$ (i.e.~$k$ connected nodes); 
\item \emph{geodesic distance distribution}, which is defined as the shortest path length between two nodes (or infinity if there is no such path); and
\item 
\emph{edge-wise shared partners distribution}, which expresses the tendency  for tied nodes to have multiple shared partners. Specifically, the edge-wise shared partner statistic counts all nodes $i$ and $j$ that are connected and also have an edge to a third node $k$. 
\end{itemize}
The measures can be visualised via boxplots using the \texttt{bgof()} function in the \texttt{bergm} package \citep{bergm}. The function requires a number of simulated networks; we set this equal to 100 later in Section~\ref{sec:empirical}. For formal definitions of the statistics, see Supplement~A. 

\section{Empirical Analyses}\label{sec:empirical}

We now consider three applications of our approach for networks requiring actor-specific information. In Section~\ref{subsec:italy}, we first consider an undirected political network with 663 nodes, 20,049 edges, and edge-specific covariates representing the Italian Parliament, in which one node represents a member of parliament.    In Section~\ref{subsec:weapons}, we consider a directed network of international arms trading in 2016 involving  163 countries, each represented by a node, and their export and import information. Previous methods typically cannot deal with all present challenges, i.e.~the need of nodal effects, nodal covariates, and the fact that the network is directed. Finally, in Section~\ref{subsec:facebook}, we illustrate the scalability of our VI method along a large undirected network with 4,039 nodes from Facebook. Owing to its size, previous work has mostly analysed subsets of this database. Further material can be found in Supplement E.

\subsection{Political collaboration network}\label{sec:parliament}
\label{subsec:italy}
We consider data from 2008--2013, which determines the last legislative period with Silvio Berlusconi as Prime Minister, though he resigned in 2011 before the end of the election period. The 663 nodes in the network are the members of parliament during the time window and edges  define the number of co-sponsorships of a law proposal.

\paragraph*{Model specification} We look at the binary network with an edge representing at least one co-sponsorship. 
As network statistics we include the following: 
\begin{itemize}
\item $s_{\mbox{\scriptsize{edge}}}(y)$,  the number of edges in the network;
\item $s_{\mbox{\scriptsize{2-star}}}(y)$, the number of 2-stars, which measures whether co-sponsorship is driven by the number of co-authorships one of the members of parliament has;
\item $s_{\mbox{\scriptsize{gwesp}}}(y)$, the number of co-sponsorship-wise shared members of parliament in a geometrically weighted form. This statistic stabilizes the estimation.
\end{itemize}
Moreover, as exogenous quantities we include
\begin{itemize}
    \item $s_{\mbox{\scriptsize{male}}}(y) = \sum_{i=1}^N \sum_{j>i} y_{ij}  \mathcal{I}\{\mbox{male}_{i} = \mbox{male}_{j}\}$,  which questions whether males are more likely to co-sponsor a proposal, i.e.\ male homophily;
    \item $s_{\mbox{\scriptsize{female}}}(y)$ being the same for females
    \item $s_{\mbox{\scriptsize{party}}}(y) = \sum_{i=1}^N \sum_{j>i} y_{ij}  \mathcal{I}\{\mbox{party}_{i} = \mbox{party}_{j}\}$ as indicator whether co-sponsors belong to the same party, and finally 
    \item $s_{\mbox{\scriptsize{age}}}(y) = \sum_{i=1}^N \sum_{j>i} y_{ij}  |\mbox{{age}}_i - \mbox{{age}}_j|$,
which quantifies age homophily.
\end{itemize}

We start with a model containing the network statistics only, denoted by $\mathcal{M}_1$. Next, we augment this model by the respective edge-specific covariates, denoted by $\mathcal{M}_2$. Finally, both models are further augmented by nodal random effects, which we label $\mathcal{M}_1^*$ and $\mathcal{M}_2^*$. Computing times for 1,000 VI iterations were 3.7, 5, 4, and 5.3 minutes for models $\mathcal{M}_1$, $\mathcal{M}_1^\ast$, $\mathcal{M}_2$, and $\mathcal{M}_2^\ast$, respectively.

\begin{figure}[htbp]
\caption{Political  collaboration network.}
    \centering
    \includegraphics[width=0.48\textwidth]{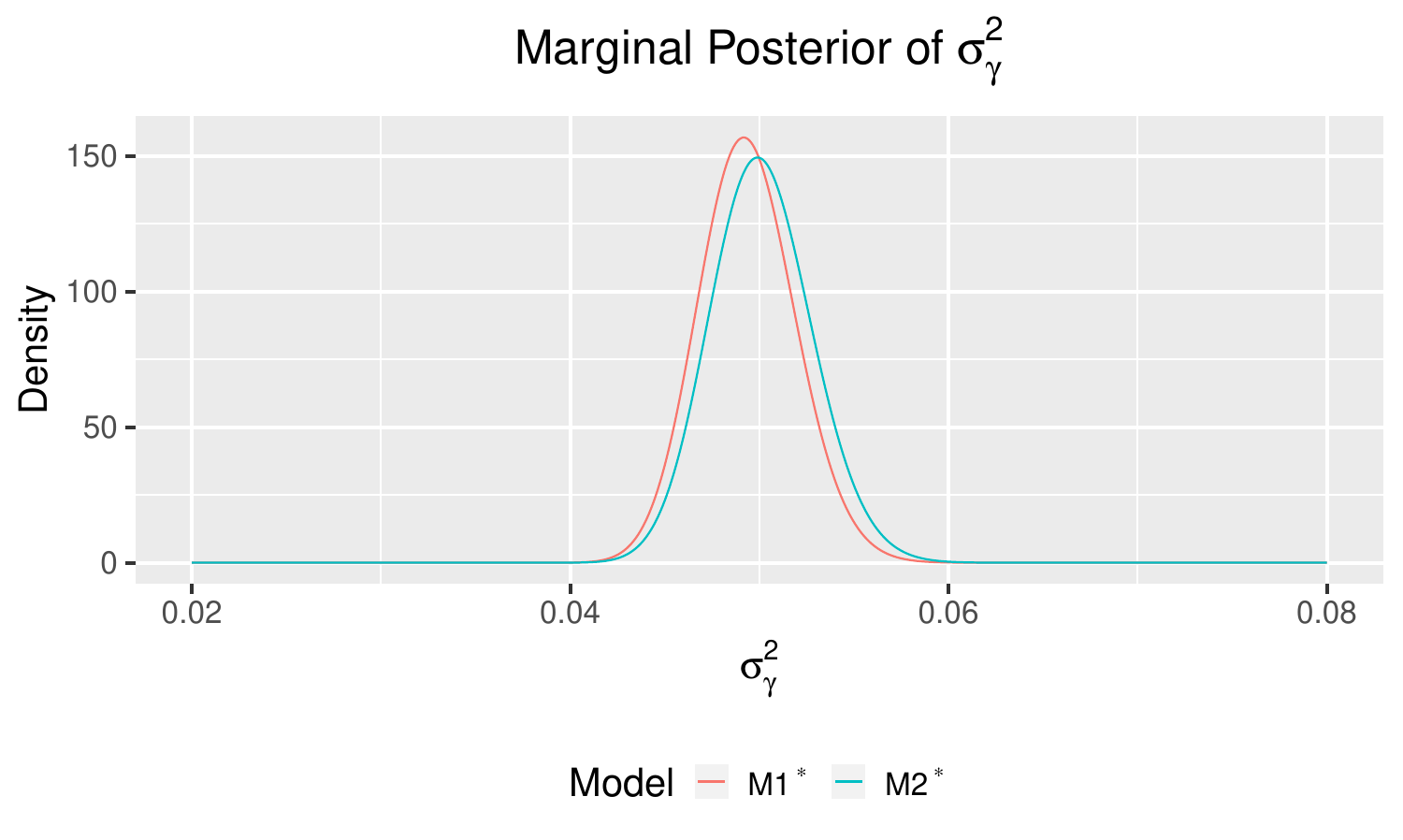}
    \caption*{Marginal posterior densities of variance parameters $\sigma_\gamma^2$ in $\mathcal{M}_1^*$ (excluding nodal covariates) and $\mathcal{M}_2^*$ (including nodal covariates).}
    \label{fig:parliament:densities}
\end{figure}

\paragraph*{Results}
As Table \ref{tab:bf_par} depicts, models containing nodal effects outperform the respective models without. In addition, the model with covariates outperforms the one without covariates. BGOFs  presented in the Supplement, Figure~E.2 confirm this observation. However, looking at posterior explained variances comparing models $\mathcal{M}_2^\ast$ and $\mathcal{M}_1^\ast$, we obtain 
    $R_{\gamma} = 2.6\mathrm{e}{-09}$. Hence, the covariates cannot explain heterogeneity as can be seen in Figure~\ref{fig:parliament:densities}.
\begin{table}[htbp]
    \caption {\small Political collaboration network.}
\renewcommand{\arraystretch}{0.4}\small
    \centering
    \begin{tabular}{c|ccc}
    \hline\hline\\
       Comp. models  & $\mathcal{M}_1^* \mid \mid \mathcal{M}_1$ & $\mathcal{M}_2^* \mid \mid \mathcal{M}_2$ & $\mathcal{M}_2^* \mid \mid \mathcal{M}_1^*$ \\\\\hline\\
log(BF) & 263.2 & 199 & 671,868.7 \\\\ 
\hline\hline
    \end{tabular}
    \label{tab:bf_par}
    \\
    \centering{\footnotesize{Log-Bayes factors for competing models.}}
\end{table}
Table~\ref{tab:coef_par} lists posterior mean estimates (together with 95\% corrected credible intervals) of network statistics for all four models. 
We find strong homophily for the party, which indeed is not surprising. Additionally, we  obtain a strong and significant gender homophily.
\begin{table*}[h]
    \caption{Political  collaboration network.}
\renewcommand{\arraystretch}{0.4}\small
    \centering
    \begin{tabular}{c|cccc}
    \hline\hline\\
       Statistics/covariate  & $\mathcal{M}_1$ & $\mathcal{M}_1^*$ & $\mathcal{M}_2$ & $\mathcal{M}_2^*$ \\\\\hline\\
$s_{\mbox{\scriptsize{edge}}}(y)$ & \begin{tabular}[x]{@{}c@{}} \textbf{-6.253} \\(-6.284, -6.221) \end{tabular} & & \begin{tabular}[x]{@{}c@{}} \textbf{-6.754} \\(-6.758, -6.750) \end{tabular} & \\\\
$s_{\mbox{\scriptsize{2-star}}}(y)$ & \begin{tabular}[x]{@{}c@{}} {$1.3\mathrm{e}{-04}$} \\($7.2\mathrm{e}{-06}$, $2.6\mathrm{e}{-04}$) \end{tabular} & \begin{tabular}[x]{@{}c@{}} ${-1.8\mathrm{e}{-04}}$ \\($-2\mathrm{e}{-04}$, $-1.6\mathrm{e}{-04}$) \end{tabular} & \begin{tabular}[x]{@{}c@{}} \textbf{0.003} \\(0.0027, 0.032) \end{tabular} & \begin{tabular}[x]{@{}c@{}} \textbf{0.003} \\ (0.0025, 0.0033) \end{tabular} \\\\ 
$s_{\mbox{\scriptsize{gwesp}}}(y)$ & \begin{tabular}[x]{@{}c@{}} \textbf{0.242} \\(0.239, 0.245) \end{tabular} & \begin{tabular}[x]{@{}c@{}} \textbf{0.236} \\(0.236, 0.236) \end{tabular} & \begin{tabular}[x]{@{}c@{}} \textbf{0.214} \\(0.214, 0.215) \end{tabular} & \begin{tabular}[x]{@{}c@{}} \textbf{0.211} \\(0.211, 0.211) \end{tabular} \\\\ 
$s_{\mbox{\scriptsize{male}}}(y)$ & & & \begin{tabular}[x]{@{}c@{}} \textbf{0.193} \\(0.184, 0.203) \end{tabular} & \begin{tabular}[x]{@{}c@{}} \textbf{0.161} \\(0.154, 0.168) \end{tabular} \\\\
$s_{\mbox{\scriptsize{female}}}(y)$ & & & \begin{tabular}[x]{@{}c@{}} \textbf{0.277} \\(0.276, 0.277) \end{tabular} & \begin{tabular}[x]{@{}c@{}} {0.223} \\(0.219, 0.227) \end{tabular} \\\\
$s_{\mbox{\scriptsize{party}}}(y)$ & & & \begin{tabular}[x]{@{}c@{}} \textbf{0.799} \\(0.798, 0.801\end{tabular} & \begin{tabular}[x]{@{}c@{}} \textbf{0.771} \\(0.765, 0.778) \end{tabular} \\\\
$s_{\mbox{\scriptsize{age}}}(y)$ & & & \begin{tabular}[x]{@{}c@{}} {0.001} \\(0.000, 0.002) \end{tabular} & \begin{tabular}[x]{@{}c@{}} {-0.001} \\(-0.002, 0.001) \end{tabular} \\\\
\hline\hline
    \end{tabular}
    \\
       \centering{\footnotesize Posterior mean estimates together with 95\% corrected  credible intervals of coefficients of network statistics and covariates for models $\mathcal{M}_1$, $\mathcal{M}_1^*$, $\mathcal{M}_2$, $\mathcal{M}_2^*$. Effects not containing the zero within the respective credible interval are highlighted in bold.}
    \label{tab:coef_par}
\end{table*}


\subsection{International arms trading}\label{subsec:weapons}

With our second example, we aim  to understand why (and when) countries trade arms. We use ERGMs to shed light on this question by considering international arms trading in 2016 as a directed network consisting of 163 countries (nodes) and 407 edges, where a directed edge refers to recorded arms trading in that year. 
The data were taken from the Stockholm International Peace Research Institute (www.sipri.org). Economically as well as politically, it is of interest to understand the driving forces behind arms trading.  Particular interest lies in identifying which covariates have a direct influence on the probability that one country exports arms to another.    Hence, in addition to network statistics, it is of interest to include  nodal as well as bi-nodal information. \citet{PerNeu2010} argue that there is a time delay between the date arms are ordered and their delivery date. Therefore, all exogenous  covariates are based on a two-year lag. 

\paragraph*{Model specification} With respect to including relevant covariates, we  follow previous work \citep[see e.g.][]{doi:10.1177/0022002718801965,Lebacher-etal:2021}; using the (log) gross domestic product for the receiver as well as for the sender $(\mbox{lgdp}_{\mbox{\scriptsize{in}}}/\mbox{lgpd}_{\mbox{\scriptsize{out}}})$, and the logarithmic military expenditure  ({lmilex}) for the receiving country. Furthermore, we account for information on alliances between countries ({alliance}) and the regime dissimilarity between two partners by measuring the absolute difference between the country's so-called polity score ({polity}$\in (-10,10)$). In addition to these exogenous quantities, we include, besides the number of edges, the following endogenous network statistics:
\begin{itemize}
    \item $s_{\mbox{\scriptsize{mutual}}}(y) $ , which counts how many reciprocal relations exist in a network, i.e. countries which mutually trade with each other.
    \item $s_{\mbox{\scriptsize{2-star-out}}}(y)$, which counts 2-star out constellations, i.e.\ one country exporting to two different  countries; 
    \item $s_{\mbox{\scriptsize{2-star-in}}}(y)$, which counts 2-star in constellations, i.e.\ one country  importing from two different  countries; 
        \item $s_{\mbox{\scriptsize{transitivity}}}(y)$, which counts constellations according to ``the trade partner of my trade partner is my trade partner'', i.e., triangles of the form that country $i$ sells weapons to countries $j$ and $k$ which themselves trade in that $j$ sells weapons to $k$.
\end{itemize}  
We start with a model containing the network statistics only, i.e. 
\begin{align*}
\mathcal{M}_1:& \exp\big(\beta_{\mbox{\scriptsize{mutual}}}s_{\mbox{\scriptsize{mutual}}}(y)+ \beta_{\mbox{\scriptsize{2-star-out}}}s_{\mbox{\scriptsize{2-star-out}}}(y)\\&\quad+\beta_{\mbox{\scriptsize{2-star-in}}}s_{\mbox{\scriptsize{2-star-in}}}(y)+\beta_{\mbox{\scriptsize{transitivity}}}s_{\mbox{\scriptsize{transitivity}}}(y)\big)
\end{align*}
and contrast it against a model with node-specific heterogeneity, i.e.
\begin{align*}
\mathcal{M}_1^\ast:& \exp\big(\beta_{\mbox{\scriptsize{mutual}}}s_{\mbox{\scriptsize{mutual}}}(y)+ \beta_{\mbox{\scriptsize{2-star-out}}}s_{\mbox{\scriptsize{2-star-out}}}(y)\\&\quad+\beta_{\mbox{\scriptsize{2-star-in}}}s_{\mbox{\scriptsize{2-star-in}}}(y)+\beta_{\mbox{\scriptsize{transitivity}}}s_{\mbox{\scriptsize{transitivity}}}(y)+t(y)^\top\gamma\big).
\end{align*}
The models are identical, despite the $2N=322$ random effects for receiver and sender that are not included in $\mathcal{M}_1$. To demonstrate that exogenous variables can explain some of the unobserved heterogeneity, we contrast $\mathcal{M}_1$ and $\mathcal{M}_1^\ast$ to the corresponding models including covariates, i.e.
\begin{align*}
\mathcal{M}_2:& \exp\big(\beta_{\mbox{\scriptsize{\text{lgdp\textsubscript{out}}}}}s_{\mbox{\scriptsize{\text{lgdp\textsubscript{out}}}}}(y)+\beta_{\mbox{\scriptsize{\text{gdp\textsubscript{in}}}}}s_{\mbox{\scriptsize{\text{gdp\textsubscript{in}}}}}(y)\\&\quad+\beta_{\mbox{\scriptsize{\text{lmilex\textsubscript{in}}}}}s_{\mbox{\scriptsize{\text{lmilex\textsubscript{in}}}}}(y)+\beta_{\mbox{\scriptsize{alliance}}}s_{\mbox{\scriptsize{alliance}}}(y)\\&\quad+
\beta_{\mbox{\scriptsize{mutual}}}s_{\mbox{\scriptsize{mutual}}}(y)+ \beta_{\mbox{\scriptsize{2-star-out}}}s_{\mbox{\scriptsize{2-star-out}}}(y)\\&\quad+\beta_{\mbox{\scriptsize{2-star-in}}}s_{\mbox{\scriptsize{2-star-in}}}(y)+\beta_{\mbox{\scriptsize{transitivity}}}s_{\mbox{\scriptsize{transitivity}}}(y\big)\\
\mathcal{M}_2^\ast:&\exp\big(\beta_{\mbox{\scriptsize{\text{lgdp\textsubscript{out}}}}}s_{\mbox{\scriptsize{\text{lgdp\textsubscript{out}}}}}(y)+\beta_{\mbox{\scriptsize{\text{gdp\textsubscript{in}}}}}s_{\mbox{\scriptsize{\text{gdp\textsubscript{in}}}}}(y)\\&\quad+\beta_{\mbox{\scriptsize{\text{lmilex\textsubscript{in}}}}}s_{\mbox{\scriptsize{\text{lmilex\textsubscript{in}}}}}(y)+\beta_{\mbox{\scriptsize{alliance}}}s_{\mbox{\scriptsize{alliance}}}(y)\\&\quad+
\beta_{\mbox{\scriptsize{mutual}}}s_{\mbox{\scriptsize{mutual}}}(y)+ \beta_{\mbox{\scriptsize{2-star-out}}}s_{\mbox{\scriptsize{2-star-out}}}(y)\\&\quad+\beta_{\mbox{\scriptsize{2-star-in}}}s_{\mbox{\scriptsize{2-star-in}}}(y)+\beta_{\mbox{\scriptsize{transitivity}}}s_{\mbox{\scriptsize{transitivity}}}(y)+t(y)^\top\gamma\big).
\end{align*}
Computing times for 1,000 VI iterations were 0.9, 1.5,  1.3, and 1.6 minutes for models $\mathcal{M}_1$, $\mathcal{M}_1^\ast$, $\mathcal{M}_2$, and $\mathcal{M}_2^\ast$, respectively.

\begin{figure*}[htbp]
\caption{International arms trading.}
    \centering
    \includegraphics[width=0.8\textwidth]{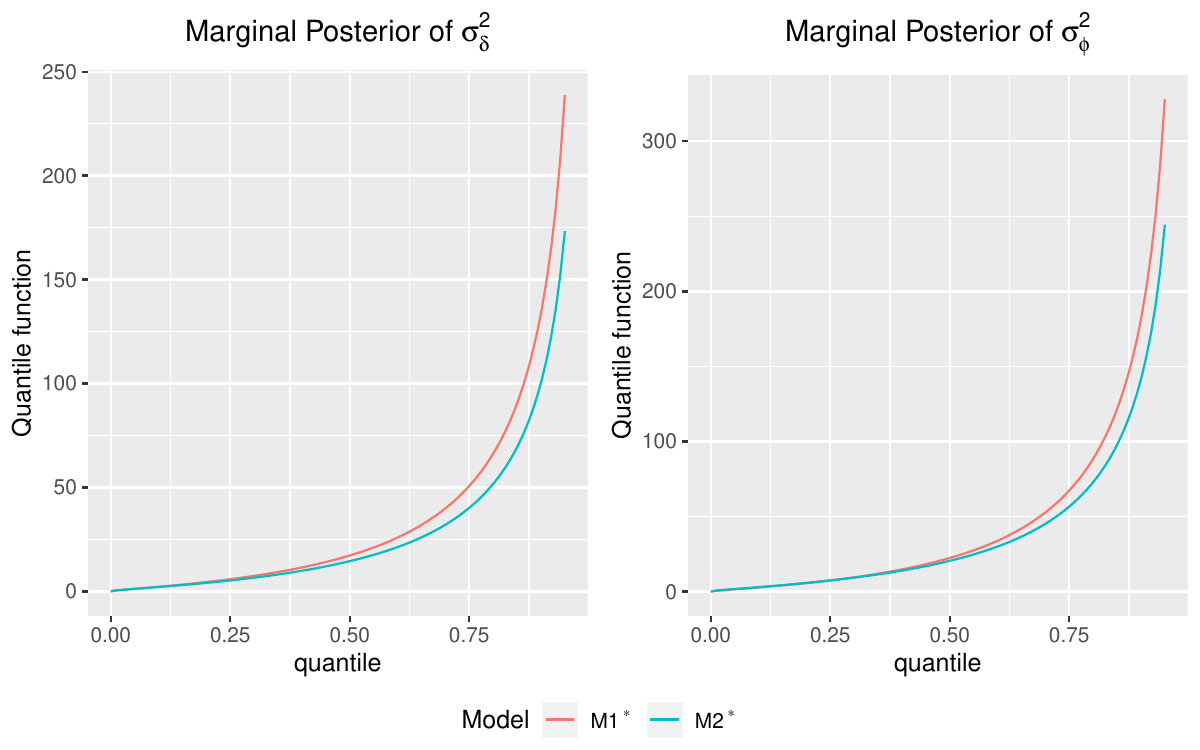}
    \caption*{Marginal posterior quantile functions of variance parameters in $\mathcal{M}_1^*$ (excluding nodal covariates) and $\mathcal{M}_2^*$ (including nodal covariates).}
    \label{fig:waffenhandel:qf}
\end{figure*}

\paragraph*{Results}
The posterior explained variances comparing models $\mathcal{M}_2^\ast$ and $\mathcal{M}_1^\ast$ are 
    $R_{\delta} = 0.55$ and $R_{\phi} = 0.52$ for the sender and receiver, respectively. Quantile functions of marginal posteriors of the variance parameters are shown in Figure~\ref{fig:waffenhandel:qf}. In addition, the respective posterior expectations of models  $\mathcal{M}_1^\ast$  vs $\mathcal{M}_2^\ast$ are 61.8 vs 45.2 and 84.8 vs 63.8  for the sender and receiver, respectively. These measures demonstrate in general that the included covariates can contribute to the reduction of unobserved heterogeneity for both exports and imports.
Moreover, the BGOFs in Figure~E.3 in the Supplement~illustrate that retaining the nodal random effects in addition to the covariates improves the model fit, such that  $\mathcal{M}_2^\ast$ is preferred  over model $\mathcal{M}_2$ without nodal effects and also over $\mathcal{M}_1^\ast$ without covariates. 

Table~\ref{tab:waffenhandel} lists the posterior mean estimates (together with 95\% corrected credible intervals) of network statistics for all four models. We see a strong negative effect for mutual trade. In other words, it is not likely that countries sell weapons to a country they buy weapons from. Transitivity is positive so that arms trading between two importing countries of the same importer  is likely. Moreover, 2-in-star is negative, which implies that countries prefer to import from only a few countries. In contrast, 2-out-star is positive, which demonstrates that countries exporting arms  tend to do so to multiple countries. As would be expected, military expenditure, GDP,  as well as the indicator as to whether the countries share an alliance have a positive effect, i.e.,\ promote the trading of weapons. The difference in polity score is not significant, and slightly negative when allowing for node heterogeneity. 
\begin{table*}[htbp]
    \caption {International arms trading.}
\renewcommand{\arraystretch}{0.4}\small
    \centering
    \begin{tabular}{c|ccccc}
    \hline\hline\\
       Statistics/covariate  & $\mathcal{M}_1$ & $\mathcal{M}_1^*$ & $\mathcal{M}_2$ & $\mathcal{M}_2^*$ \\\\\hline\\
$s_{\mbox{\scriptsize{edge}}}(y)$ & \begin{tabular}[x]{@{}c@{}}\textbf{-5.836} \\(-5.847, -5.824) \end{tabular} & \begin{tabular}[x]{@{}c@{}}\textbf{-4.655} \\(-4.673, -4.637) \end{tabular} & \begin{tabular}[x]{@{}c@{}}\textbf{-5.958} \\(-5.958, -5.957) \end{tabular} & \begin{tabular}[x]{@{}c@{}}\textbf{-5.888} \\(-5.889, -5.886) \end{tabular} \\\\
$s_{\mbox{\scriptsize{mutual}}}(y)$ & \begin{tabular}[x]{@{}c@{}}\textbf{-0.762} \\(-0.762, -0.762) \end{tabular} & \begin{tabular}[x]{@{}c@{}}\textbf{-0.734} \\(-0.734, -0.734) \end{tabular} & \begin{tabular}[x]{@{}c@{}}\textbf{-0.652} \\(-0.652, -0.652) \end{tabular} & \begin{tabular}[x]{@{}c@{}}\textbf{-0.600} \\(-0.601, -0.599) \end{tabular} \\\\
$s_{\mbox{\scriptsize{transitive}}}(y)$ & \begin{tabular}[x]{@{}c@{}}\textbf{0.490} \\(0.490, 0.491) \end{tabular} & \begin{tabular}[x]{@{}c@{}}\textbf{0.277} \\(0.277, 0.278) \end{tabular} & \begin{tabular}[x]{@{}c@{}}\textbf{0.409} \\(0.408, 0.409) \end{tabular} & \begin{tabular}[x]{@{}c@{}}\textbf{0.255} \\(0.254, 0.255) \end{tabular} \\\\
$s_{\mbox{\scriptsize{2-in-star}}}(y)$ & \begin{tabular}[x]{@{}c@{}}\textbf{0.098} \\(0.087, 0.109) \end{tabular} & \begin{tabular}[x]{@{}c@{}}\textbf{-0.013} \\(-0.014, -0.013) \end{tabular} & \begin{tabular}[x]{@{}c@{}}\textbf{-0.029} \\(-0.032, -0.026) \end{tabular} & \begin{tabular}[x]{@{}c@{}}\textbf{-0.109} \\(-0.110, -0.108) \end{tabular} \\\\
$s_{\mbox{\scriptsize{2-out-star}}}(y)$ & \begin{tabular}[x]{@{}c@{}}\textbf{0.086} \\(0.075, 0.097) \end{tabular} & \begin{tabular}[x]{@{}c@{}}\textbf{0.075} \\(0.074, 0.076) \end{tabular} & \begin{tabular}[x]{@{}c@{}}\textbf{0.054} \\(0.050, 0.057) \end{tabular} & \begin{tabular}[x]{@{}c@{}}\textbf{0.041} \\(0.040,  0.041) \end{tabular} \\\\
$lmilex$& & & \begin{tabular}[x]{@{}c@{}}\textbf{0.056} \\(0.050, 0.062) \end{tabular} & \begin{tabular}[x]{@{}c@{}}\textbf{0.030} \\(0.013, 0.046) \end{tabular} \\\\
$gdp_{out}$& & & \begin{tabular}[x]{@{}c@{}}\textbf{0.358} \\(0.348, 0.367) \end{tabular} & \begin{tabular}[x]{@{}c@{}}\textbf{0.138} \\(0.118, 0.159) \end{tabular} \\\\
$gdp_{in}$& & & \begin{tabular}[x]{@{}c@{}}\textbf{0.054} \\(0.049, 0.058) \end{tabular} & \begin{tabular}[x]{@{}c@{}}\textbf{-0.131} \\(-0.145 -0.118) \end{tabular} \\\\
$polity$& & & \begin{tabular}[x]{@{}c@{}}{-0.023} \\(-0.058, 0.011) \end{tabular} & \begin{tabular}[x]{@{}c@{}}\textbf{-0.041} \\(-0.060, -0.022) \end{tabular} \\\\
$alli$& & & \begin{tabular}[x]{@{}c@{}}\textbf{0.526} \\(0.525, 0.527) \end{tabular} & \begin{tabular}[x]{@{}c@{}}\textbf{0.372} \\(0.371, 0.374) \end{tabular} \\\\
\hline\hline
    \end{tabular}
    \\
       \centering{\footnotesize{Posterior mean estimates together with 95\% corrected  credible intervals of coefficients of network statistics and covariates for models $\mathcal{M}_1$, $\mathcal{M}_1^*$, $\mathcal{M}_2$, $\mathcal{M}_2^*$. Effects not containing the zero within the respective credible interval are highlighted in bold.}}
    \label{tab:waffenhandel}
\end{table*}

\subsection{Facebook ego network}\label{subsec:facebook}
The third example is a benchmark dataset taken from the Stanford Large Network Dataset Collection \citep[SNAP datasets;][]{LesKre2014}. The dataset provides information on friend lists from ten anonymized Facebook users called \emph{egos}.  Given the fact that the egos are connected to all their friends, the egos (and all nodes only having connections to egos) are  excluded before analysing the network, as too much importance would otherwise be attributed to those nodes; see Figure~\ref{fig:facebook}. However, most papers so far have only considered a subsample of the network due to its size \citep[e.g.,][]{ThiKau2017}. The focus here is to demonstrate that the variational approach allows us to fit and select complex models for large networks, even in the presence of heterogeneity.


\paragraph*{Model specification}
We select the model with two major aims, $i)$ to identify relevant network statistics and $ii)$ decide whether or not nodal effects improve the model fit.
For $i)$ we specifically compare the three models $\mathcal{M}_1$ to $\mathcal{M}_3$ with the following model specifications:
\begin{equation*}
\begin{aligned}
\mathcal{M}_1: & \exp(\beta_{\mbox{\scriptsize{edge}}}s_{\mbox{\scriptsize{edge}}}(y)+  \beta_{\mbox{\scriptsize{2-star}}}s_{\mbox{\scriptsize{2-star}}}(y)+\beta_{\mbox{\scriptsize{triangle}}}s_{\mbox{\scriptsize{triangle}}}(y))\\
\mathcal{M}_2: & \exp(\beta_{\mbox{\scriptsize{{edge}}}}s_{\mbox{\scriptsize{edge}}}(y)+  \beta_{\mbox{\scriptsize{2-star}}}s_{\mbox{\scriptsize{2-star}}}(y)+ \beta_{\mbox{\scriptsize{gwesp}}}s_{\mbox{\scriptsize{gwesp}}}(y)) \\
\mathcal{M}_3: & \exp(\beta_{\mbox{\scriptsize{edge}}}s_{\mbox{\scriptsize{edge}}}(y)+ \beta_{\mbox{\scriptsize{2-star}}}s_{\mbox{\scriptsize{2-star}}}(y) + \beta_{\mbox{\scriptsize{triangle}}}s_{\mbox{\scriptsize{triangle}}}(y)\\&\quad+ \beta_{\mbox{\scriptsize{gwesp}}}s_{\mbox{\scriptsize{gwesp}}}(y)).
\end{aligned}
\end{equation*}
For $ii)$ we include nodal random effects in each model and denote these as $\mathcal{M}_1^\ast$ to $\mathcal{M}_3^\ast$. 

\paragraph*{Results}
The computing times for 1,000 VI iterations were 14, 19.1, and 37.7 minutes for $\mathcal{M}_1$ to $\mathcal{M}_3$ and 142.5, 156.3, and 162.2 minutes for $\mathcal{M}_1^\ast$ to $\mathcal{M}_3^\ast$, respectively.
\begin{table*}[htbp] \renewcommand{\arraystretch}{1.5}
\caption{ Facebook egos.}\label{tab:facebook:BF}
\centering\begin{tabular}{c|cccccc}\hline\hline
Comp.~models & $\mathcal{M}1^\ast||\mathcal{M}1$ & $\mathcal{M}_2^\ast||\mathcal{M}_2$ & $\mathcal{M}_3^\ast||\mathcal{M}_3$ & $\mathcal{M}_3^\ast||\mathcal{M}_1^\ast$ & $\mathcal{M}_2^\ast||\mathcal{M}_1^\ast$ & $\mathcal{M}_3^\ast||\mathcal{M}_2^\ast$ \\ 
\hline
log(BF)  & $1.2 \times 10^3$ &
 $46 \times 10^3$ & $51 \times 10^3$  &  $244 \times 10^6$  & $ 16 \times 10^6$  &  $-208 \times 10^6$
\\
\hline\hline
\end{tabular}
\\
\centering{\footnotesize{Log-Bayes factors for competing models.}}
\end{table*}
As can be seen in Table~\ref{tab:facebook:BF}, models containing nodal effects outperform the respective ones without.   $\mathcal{M}_2^\ast$ and $\mathcal{M}_3^\ast$ outperform $\mathcal{M}_1^\ast$ whereas $\mathcal{M}_2^\ast$ is preferred over $\mathcal{M}_3^\ast$ in terms of BFs. This observation is confirmed by the BGOFs  in the Supplement,~Figure~E.5. 

\begin{table*}[ht] 
\caption{Facebook egos. }\renewcommand{\arraystretch}{0.4}\small
\centering
\begin{tabular}{c|ccccc}
\hline\hline\\
\diagbox{Model}{Statistic} & $s_{\mbox{\scriptsize{edge}}}(y)$ & $s_{\mbox{\scriptsize{triangles}}}(y)$ & $s_{\mbox{\scriptsize{2-star}}}(y)$ & $s_{\mbox{\scriptsize{gwesp}}}(y)$ & $\gamma$ \\\\
 \hline\hline\\
$\mathcal{M}_1$  & \textbf{-6.157} & \textbf{0.581 }& \textbf{-0.016} & & No\\
  & (-6.159,-6.156) & (0.580,0.581) & (-0.017, -0.015) & &\\ \\
$\mathcal{M}_1^\ast$  & & \textbf{0.594} & \textbf{-0.0246}  & & Yes\\
 & & (0.559,0.629) & (-0.02464,-0.0245) & & \\ \\
$\mathcal{M}_2$  & \textbf{-7.217} & & \textbf{-0.012} & \textbf{0.554} & No \\
 & (-7.240, -7.193) & & (-0.021, -0.003) & (0.384,0.724) &  \\\\
$\mathcal{M}_2^\ast$  &  & & \textbf{-0.018}  & \textbf{0.586 }& Yes \\
 &  & & (-0.0188,-0.0174) &  (0.503.0.668) & \\ \\
$\mathcal{M}_3$  & \textbf{-6.955} & 0.026 & \textbf{-0.0165} & \textbf{0.524} & No\\
 & (-6.978,-6.932) & (-0.011,0.063) & (-0.023, -0.010) & (0.412, 0.636) & \\\\
$\mathcal{M}_3^\ast$  &  &\textbf{ 0.067} &\textbf{ -0.035} & \textbf{0.550 }& Yes\\
 &  & (0.059,0.075) & (-0.036,-0.035) & (0.527,0.573) & \\\\
 \hline\hline
\end{tabular}\vspace{0.2cm}\caption*{Posterior mean estimates together with 95\% corrected credible intervals of coefficients of network statistics for all models. Effects not containing the zero within the credible interval are highlighted in bold.
}\label{tab:fab}
\end{table*}
Table~\ref{tab:fab} lists the posterior mean estimates (together with 95\% corrected credible intervals) of network statistics for all six models. A deeper discussion on interpreting effects of network statistics in general is provided in \cite{SniPatRobHan2006}. Here, we briefly highlight some findings in the best model $\mathcal{M}_2^\ast$ (4th row in Table~\ref{tab:fab}).   
 The effect $\beta_{\mbox{\scriptsize{2-star}}}$ is (statistically significantly) negative, which indicates that the likelihood of a tie decreases if an additional 2-star exists in the network, meaning one node has two friends in common. This is a partial effect, given the  effect $\beta_{\mbox{\scriptsize{gwesp}}}$, which is  (statistically significantly) positive meaning that, overall, there is a tendency to increase the number of joint partners. However, as nodal heterogeneity is required following the BF-based model selection, the actors in the network are too heterogeneous to describe Facebook friendships just by the selected statistics. 

\section{Conclusion}\label{sec:conclusion}
In this paper, we have developed variational inference for the analysis of large scale networks with nodal heterogeneity. Our approach comes with a number of merits and solutions in comparison to existing methods. Firstly, it allows the incorporation of nodal heterogeneity in the network, both in undirected as well as in directed networks. Secondly, while previous Bayesian network models  are unsuitable for large networks owing to their numerical complexity, the variational approach pushes this limit, allowing us to analyse networks with thousands of nodes even when node heterogeneity is included.  Thirdly, we contribute to model selection and model validation, utilizing the information obtained through the (approximate) posterior distributions. In particular, comparing models with and without nodal heterogeneity effects provides a novel framework for assessing how much external node or edge-specific covariates contribute to explaining the existence of edges. This is demonstrated in a directed network on arms trading. Lastly, we propose a simple correction step, so that the posterior standard errors are closer to the true ones. 

Overall, we demonstrate the great potential of variational methods in the context of network analyses with a large number of edges and nodal heterogeneity. In the future, it would be of interest to leverage this potential further to networks that are beyond the scope of this paper, e.g.~for dynamic \citep{LebThuKau2021} or multi-edge \citep{BraCasNanSch2019} networks. In the context of such networks, other choices of variational approximation may also be investigated, such as methods based on implicit copulas \citep{SmiLoa2022} or mixture models \citep{GunKohNot2021}.

\

\paragraph{Acknowledgments}
The authors would like to thank Jana Kleinemeier and Tim B\"undert for assistance in coding and evaluation of simulation results and Prof.~Michael Smith for an initial discussion on this project. Andrew Entwistle kindly provided his English proofreading expertise. We thank Cornelius Fritz and Michael Lebacher for data sharing and useful discussions on the empirical illustrations.

\paragraph{Funding}
The first author Nadja Klein was supported by the Deutsche Forschungsgemeinschaft (DFG, German Research Foundation) through the Emmy Noether grant KL 3037/1-1.

\paragraph{Supplementary Material}
{This supplement contains the employed network statistics (A), further derivations and calculations relevant to our VI approach (B--D), additional material on the two real data examples (E), and extensive simulations evaluating the accuracy and robustness of our estimator, as well as benchmarking with exact Bayesian inference in small networks (F).}


\bibliographystyle{dcu} 
\bibliography{references}       

\end{document}